\documentclass[letterpaper,twocolumn]{IEEEtran}

\IEEEoverridecommandlockouts
\pdfoutput=1

\usepackage[T1]{fontenc}
\usepackage[latin1]{inputenc}
 \usepackage{verbatim,multirow}
 \usepackage{float}
 \usepackage{amsmath}
 \usepackage{graphicx}

 \usepackage{amssymb}
\usepackage{amsthm}
 \usepackage{calc}

 \makeatletter

\usepackage{gastex}

 \floatstyle{ruled}
 \newfloat{algorithm}{tbp}{loa}
 \floatname{algorithm}{Algorithm}

 \usepackage{cite}
 \usepackage{hhline}
 \usepackage{subfigure}
\usepackage[small]{caption}
 \usepackage{array}
 \usepackage{acronym}
\usepackage{color}
\usepackage{algorithm}
\usepackage{algorithmic}
\usepackage{float}
\usepackage{psfrag}
\theoremstyle{remark}

\theoremstyle{definition}

 \newenvironment{definition}[1][Definition]{\begin{trivlist}
 \item[\hskip \labelsep {\bfseries #1}]}{\end{trivlist}}

 \makeatletter \newcommand \listoftodos{\chapter{Todo list} \@starttoc{todocontents}}
 \newcommand\l@todo[2]
     {\par\noindent pg \textit{#2}: \parbox{10cm}{#1}\par} \makeatother

\newcommand{\etal}{\textit{et al.}~}

\acrodef{RV}{random variable}

\acrodef{i.i.d.}{independent, identically distributed}

\acrodef{PDF}{probability distribution function}
\acrodef{PMF}{probability mass function}
\acrodef{CDF}{cumulative distribution function}
\acrodef{HF}{high frequency}
\acrodef{ch.f.}{characteristic function}

\acrodef{AWGN}{additive white gaussian noise}

\acrodef{SNR}{signal-to-noise ratio}

\acrodef{LRT}{likelihood ratio test}

\acrodef{GLRT}{generalized likelihood ratio test}

\acrodef{CRLB}{Cram\'{e}r-Rao lower bound}

\acrodef{CRB}{Cram\'{e}r-Rao bound}

\acrodef{ZZLB}{Ziv-Zakai lower bound}

\acrodef{ZZB}{Ziv-Zakai bound}

\acrodef{LOS}{line-of-sight}

\acrodef{NLOS}{non-line-of-sight}

\acrodef{GDOP}{geometric dilution of precision}

\acrodef{GPS}{Global Positioning System}

\acrodef{FIM}{Fisher information matrix}

\acrodef{PEB}{position error bound}

\acrodef{TOA}{time-of-arrival}

\acrodef{TOF}{time-of-flight}

\acrodef{WSN}{Wireless Sensor Network}

\acrodef{MAC}{medium access control}

\acrodef{RSS}{received signal strength}

\acrodef{WAF}{wall attenuation factor}

\acrodef{TDOA}{time difference-of-arrival}

\acrodef{RF}{radiofrequency}

\acrodef{RTT}{round-trip time}

\acrodef{AOA}{angle-of-arrival}

\acrodef{MF}{matched filter}

\acrodef{ED}{energy detector}

\acrodef{ML}{maximum likelihood}

\acrodef{MSE}{mean square error}

\acrodef{RMSE}{root mean square error}

\acrodef{p.p.m.}{part-per-million}

\acrodef{ACK}{acknowledge}

\acrodef{UWB}{ultrawide bandwidth}

\acrodef{TNR}{threshold-to-noise ratio}

\acrodef{NLOS}{non line-of-sight}

\acrodef{LOS}{line-of-sight}

\acrodef{LS}{least squares}

\acrodef{IR-UWB}{impulse radio UWB}

\acrodef{FCC}{Federal Communications Commission}

\acrodef{TH}{time-hopping}

\acrodef{PPM}{pulse position modulation}

\acrodef{MUI}{multi-user interference}

\acrodef{PDP}{power delay profile}

\acrodef{BPZF}{band-pass zonal filter}

\acrodef{SIR}{signal-to-interference ratio}

\acrodef{RFID}{radio frequency identification}

\acrodef{WPAN}{wireless personal area networks}

\acrodef{WWB}{Weiss-Weinstein bound}

\acrodef{DP}{direct path}

\acrodef{MF}{matched filter}

\acrodef{MMSE}{minimum-mean-square-error}

\acrodef{SBS}{serial backward search}

\acrodef{NBI}{narrowband interference}

\acrodef{WBI}{wideband interference}

\acrodef{INR}{interference-to-noise ratio}

\acrodef{CR}{channel response}

\acrodef{CIR}{channel impulse response}

\acrodef{LRT}{likelihood ratio test}

\acrodef{MUR}{Multistatic RADAR}

\acrodef{SPA}{sum-product algorithm}

\acrodef{MANET}{mobile ad hoc network}

\acrodef{BP}{belief propagation}
\acrodef{GBP}{generalized belief propagation}
\acrodef{Net-FG}{network factor graph}
\acrodef{Net-BP}{network belief propagation}
\acrodef{Net-RG}{network region graph}
\acrodef{NRG}{network region graph}
\acrodef{CNRG}{clustered network region graph}
\acrodef{SP}{survey propagation}
\acrodef{CLS}{cooperative least-squares}
\acrodef{FE}{free energy}

\acrodef{MST}{minimum spanning tree}
\acrodef{MN}{maxent-normal}
\acrodef{CDS}{connected dominating set}
\acrodef{CDG}{connected dominating graph}
\acrodef{MDS}{minimum dominating set}
\acrodef{MCDS}{minimum connected dominating set}
\acrodef{DS}{dominating set}
\acrodef{CH}{cluster head}
\acrodef{DN}{dominating node}
\acrodef{RG}{region graph}
\acrodef{FG}{factor graph}
\acrodef{SPAWN}{Sum-product algorithm over a wireless network}
\acrodef{NBP}{non-parametric belief propagation}
\acrodef{IS}{importance sampling}
\acrodef{MCMC}{Monte Carlo Markov Chain}
\acrodef{RS}{rejection sampling}
\acrodef{MN}{maxent-normal}
\acrodef{TRP}{tree-based reparameterization}
\acrodef{QOS}{quality of service}

\acrodef{MR}{multi-resolution}
\acrodef{SR}{single-resolution}
\acrodef{MD}{multi-description}
\acrodef{UCS}{uncoded storage}
\acrodef{URS}{uncoded resolution-aware storage}
\acrodef{CRS}{coded resolution-aware storage}

\acrodef{NPV}{net present value}
\acrodef{PV}{present value}
\acrodef{CF}{cash flow}
\acrodef{DCF}{discounted cash flow}
\acrodef{WACC}{weighted average cost of capital}

\acrodef{CDN}{content distribution network}
\acrodef{NC}{network coding}
\acrodef{DC}{data center}
\acrodef{PUE}{power usage effectiveness}
\acrodef{RLC}{random linear coding}
\acrodef{ACPI}{advanced configuration and power interface}
\acrodef{RLNC}{random linear network coding}
\acrodef{SAN}{storage area network}
\acrodef{NAS}{network attached storage}
\acrodef{NCS}{network coded storage}
\acrodef{AWS}{Amazon Web Services}
\acrodef{HD}{high definition}
\acrodef{HLS}{HTTP Live Streaming}
\acrodef{NCC}{no coefficient-cycling}
\acrodef{CC}{coefficient-cycling}
\acrodef{HDD}{hard disk drive}
\acrodef{SSD}{solid state drive}
\acrodef{TPP}{traveling purchaser problem}
\acrodef{TSP}{traveling salesman problem}
\acrodef{MS}{modern systems}

\author{Ulric J.~Ferner, Tong Wang, Muriel M\'{e}dard and Emina Soljanin\thanks{This material is based upon
    work supported by  the Jonathan Whitney  MIT
    fellowship, Alcatel-Lucent under award \#4800484399, the Air Force Office of Scientific Research under award \#FA9550-09-1-0196, Georgia Institute of Technology under award \#RA306-S1, and France Telecom S.A.~under award \#0050012310-A100. U.~J.~Ferner and M.~M\'{e}dard are with the Research Laboratory for Electronics,  Massachusetts Institute of Technology, Room 36-512, 77 Massachusetts Avenue, Cambridge,
    MA 02139 (e-mail: {uferner,medard}@mit.edu). T.~Wang is at the Prediction Analysis Laboratory, MIT, Room E62-576, 100 Main Street, Cambridge,
    MA 02142 (e-mail: tongwang@mit.edu).  E.~Soljanin is at Bell Labs, Alcatel-Lucent,
    600 Mountain Av.,  Murray Hill, NJ 07974 (e-mail: emina@research.bell-labs.com).}}
\title{Resolution-aware network coded storage}
\date{\today}

\begin{document}
\maketitle

\begin{abstract}
In this paper, we show that coding can be used in \acp{SAN} to improve various quality of service metrics under normal \ac{SAN} operating conditions, without requiring additional storage space.  For our analysis, we develop a model which captures  modern characteristics such as constrained I/O access bandwidth limitations.  Using this model, we consider two important cases: \ac{SR} and \ac{MR} systems.  For \ac{SR} systems, we use blocking probability as the quality of service metric and propose the \ac{NCS} scheme as a way to reduce blocking probability.  The \ac{NCS} scheme codes across file chunks in time,  exploiting file striping and file duplication.  Under our assumptions, we illustrate cases where \ac{SR} \ac{NCS} provides an order of magnitude savings in blocking probability.  For \ac{MR} systems, we introduce saturation probability as a quality of service metric to manage multiple user types, and we propose the \ac{URS} and \ac{CRS} schemes as ways to reduce saturation probability.  In \ac{MR} \ac{URS}, we  align our \ac{MR} layout strategy with traffic requirements.  In \ac{MR} \ac{CRS}, we code videos across \ac{MR} layers.  Under our assumptions, we illustrate that \ac{URS} can in some cases provide an order of magnitude gain in saturation probability over classic non-resolution aware systems.  Further, we illustrate that \ac{CRS} provides additional saturation probability savings over \ac{URS}.
\end{abstract}


\section{Introduction}
\label{sec:intro}

Current projections indicate that the worldwide \ac{DC} industry will require a quadrupling of
capacity by the  year 2020 \cite{ForKapKin:08}, in large part owing to rapidly growing demand for
high-definition video streaming.
Storage area networks (SANs), whose structures help determine \ac{DC} storage capacity, are designed with the goal of serving  large numbers of content consumers  concurrently, while maintaining an acceptable user experience.  When \acp{SAN} fail to meet video requests at scale, sometimes the consequences are large and video streaming providers can even be met with negative press coverage \cite{Itz:NYT12}.  

To avoid such issues one goal of \acp{SAN} is to reduce the blocking and saturation probabilities, or the probability that there is an interrupt on during consumption.  To achieve this goal, individual content is replicated on multiple drives \cite{RabSpa:B02}.  This replication increases the chance that, if a server or drive with access to target content is unavailable, another copy of the same file on a different  drive can be read instead.   Modern content replication strategies are designed to help \acp{SAN} service  millions of video requests in parallel \cite{KoJun:p06}.  

The diversity of devices used to consume video complicates video file replication requirements. The resolutions enabled by popular devices, such as smartphones, tablets, and computers, span a wide range from 360p to HD 1080p.  \acp{SAN} need to optimize for both saturation probability reduction and video resolution diversity using their limited storage.
The use of \acf{MR} codes has received increasing attention as a technique to manage video resolution diversity directly.  An \ac{MR} code is a single code which can be read at different rates, yielding reproductions at different resolutions \cite{Eff:01}.  \ac{MR} codes are typically composed of a base layer containing the lowest video resolution and refinement layers that provide additional resolution.  For instance, an H.264 or Silverlight 480p version of a video may be encoded using a 380p base layer and one or more refinement layers.  Users or their applications determine how many layers to request depending on their available communication bandwidth and video resolution preferences.  

In modern systems, different layers of \ac{MR} video are usually stored on single drives \cite{SchMarWie:07,ZhoRos:03,WeiSulBjoLut:03}.  In this paper, we propose a \acf{CRS} scheme that replaces refinement layers with pre-network coded refinement and base layers.  We are interested in the benefits of coding, and to isolate the effect of coding from the storage reallocation, we introduce the \acf{URS} scheme in which refinement layers are stored on drives different to those that store base layers. 

Decoding a refinement layer with higher resolution requires decoding its base layer and all lower refinement layers; thus, a base layer will always be in demand whenever a video of any resolution is requested.  If a system stores too few base layers, it risks the base layer drives becoming overwhelmed by user requests, causing a high probability of system saturation.  If a system instead allocates more drives for base layers at the expense of refinement layers, it risks reducing or restricting users' quality of experience.  In this manuscript, we propose a flexible approach to managing this trade-off.

This paper builds upon two prior works by the authors.  Reference \cite{FerMedSol:12} introduces blocking probability as a metric for \acp{SAN}, and details a \ac{SR} \ac{NCS} blocking probability reduction scheme.  Reference \cite{FerWanMed:13} introduces the saturation probability metric for \ac{MR} schemes, and the \ac{MR} \ac{CRS} scheme.  

The goal of this paper is to provide not only a unified view of the aforementioned \ac{SR} \cite{FerMedSol:12} and \ac{MR} schemes \cite{FerWanMed:13}, but also to contrast the  blocking and saturation probability savings of network coded storage against uncoded storage.   This paper includes:
\begin{itemize}
      \item A general \ac{SAN} model that captures the I/O access bandwidth limitations of modern systems;
    \item An analysis of \ac{SR} \ac{UCS} and \ac{NCS} in which we code across file chunks and compare their blocking probabilities;
    \item An analysis of \ac{MR} \ac{URS} and \ac{CRS} in which we code across \ac{MR} layers and compare their saturation probabilities;
    \item  For some \ac{SR} use cases, results that illustrate an order of magnitude reduction in blocking probability;
     \item  For some \ac{MR} use cases, results illustrate an order of magnitude reduction in saturation probability between classical non-resolution aware schemes and \ac{URS}.  Further savings, of yet another order of magnitude, can be found by employing \ac{CRS}.  
\end{itemize}
A toy problem showing the intuition of the schemes detailed in this paper is shown in Section \ref{sec:mainIdea} and Fig.~\ref{fig:intuition}.

We refer the reader to \cite{DimRamWuSuh:11} for a broad overview of network codes for distributed storage, especially for repair problems.  For \ac{SR} \ac{NCS} to increase the speed of data distribution and download speeds, the use of network coding primarily in peer-to-peer networks has garnered significant attention
\cite{GkaRod:05,GkaMilRod:06,ChiYeuHuaFan:06}.  In distributed storage systems, network coding has been proposed for data repair problems \cite{AceDebMedKoe:05,DimGodWaiRam:07,GerXiaSko:11}, in which drives fail and the network needs to adapt to those failed nodes by, for instance, replacing all the lost coded content at a new node.  It has also been proposed to reduce queueing delay and latency in data centers \cite{HuaPawHasRam:12,ShaLeeRam:12}, and to speed up content download \cite{JosLiuSol:12}.  
However,  to the authors' knowledge, little work has considered the application of network coding in
\acp{SAN} to achieve blocking probability improvements under normal operating conditions.

For \ac{MR} communication rather than storage schemes, related works to this manuscript on \ac{MR} codes include Kim \etal who proposed a pushback algorithm to generate network codes for single-source multicast of multi-resolution codes \cite{KimLucShiZhaMed:10}.   Soldo \etal studied a system architecture for network coding-based \ac{MR} video-streaming in wireless environments \cite{SolMarTol:10}.

The remainder of this paper is organized as follows.  Section \ref{sec:model} introduces the system model.  Section \ref{sec:analysis} details the \ac{SR} analysis and comparison, and Section \ref{sec:systematicNCS} details the \ac{MR} analysis and comparison.  Finally, Section \ref{sec:futWork} discusses findings and Section \ref{sec:conc} concludes.  See Table \ref{tab:sections} for a listing of each scheme considered in this paper its corresponding analysis section.  

\begin{table}[tb]
  \centering
  \caption{List of schemes in this paper and their corresponding analysis sections.}
  \label{tab:sections}
  \begin{tabular}{|l|l|c|}
    \hline
     No.~of & & \\
    resolutions & Scheme & Sec.\\
    \hline
 Single & Uncoded storage & \ref{sec:analysis}  \\
    Single & Network coded storage & \ref{sec:ncs-design}\\
    Multi & Classic non-resolution aware storage  & \ref{sec:systemConstraintsClassicAnal}\\
     Multi & Uncoded resolution-aware storage  & \ref{sec:results}\\
     Multi & Coded resolution-aware storage & \ref{sec:results}\\
    \hline
  \end{tabular}
\end{table}

\section{System Model}
\label{sec:model}

First, this section describes the main idea in this paper through an intuitive example.  Second, it provides the general service model and third, it describes the system model details for both the single- and multi-resolution schemes.

\subsection{An Intuitive Example}
\label{sec:mainIdea}

Consider classic circuit switched networks such as phone systems where, if the line in the network is busy for a user, their call is not queued or retired, but instead lost forever.   Under certain conditions, the blocking probability derived from the Erlang distribution, i.e., the probability of call loss in a group of circuits, can be given by the well studied Erlang-B formula from queueing theory,
\begin{align}
  P_b = \frac{ \rho^K / K! }{ \sum_{i=0}^K \rho^i/i! } \, ,
\label{eq:PbOne}
\end{align}
where $ \rho$ is the system load and $K$ is the total number of service units in the network \cite{Kle:B75}.  Consider a \ac{SAN} system with similar blocking characteristics, in which calls are replaced with user requests for file segments, and service units are replaced with drives storing content.  A \ac{SAN} is a dedicated storage area network composed of servers and drives that provides access to consolidated block-level data storage.  To reduce blocking probability, \acp{SAN} can replicate content on multiple drives to increase $K$, as per (\ref{eq:PbOne}).

\begin{figure}[tb]
  \centering
  \includegraphics[width = \linewidth]{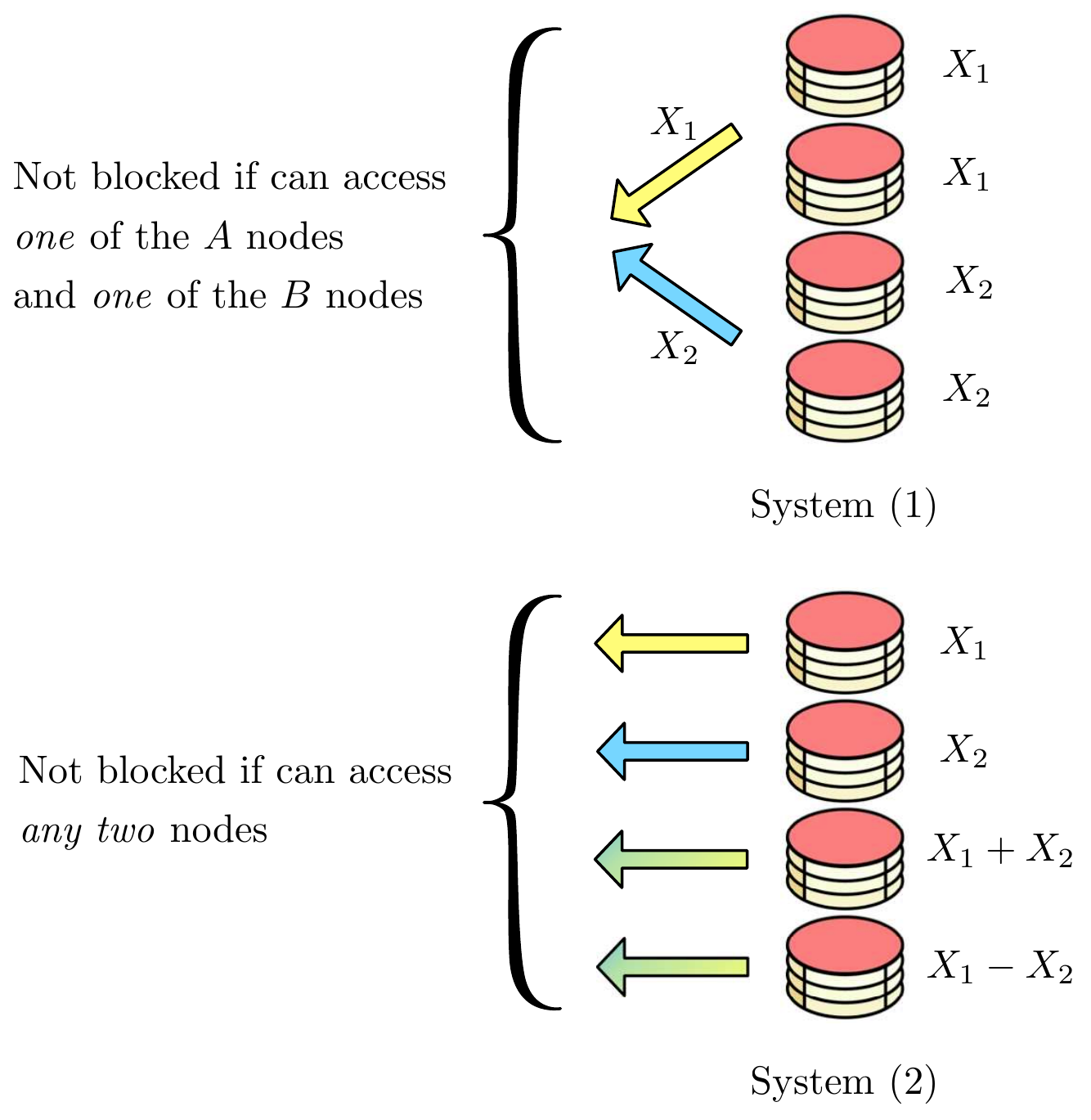}
  \caption{Intuition behind the system presented in this paper.  Consider content with two segments $X_1$ and $X_2$ and compare the following four drive systems.   System (1) is a replication system storing $ X_1, X_1, X_2, X_2$, and System (2) is a coding system storing $X_1, X_2, X_1+X_2, X_1-X_2$.  Assume a user requests both $X_1$ and $X_2$.  In System (1), the user will not be blocked if they access at least one of the top two drives storing $X_1$, and at least one of the two drives storing $X_2$.  In System (2), the user be will not be blocked and able to decode if they access \textit{any two} drives.}
  \label{fig:intuition}
\end{figure}

In this paper we consider coding file segments that are stored on different drives in the same \ac{SAN}.  Consider Fig.~\ref{fig:intuition} which depicts a content with two segments $X_1$ and $X_2$ and compares two four-drive systems.  System (1) is a replication system storing $ X_1, X_1, X_2, X_2$, and System (2) is a coding system storing $X_1, X_2, X_1+X_2, X_1-X_2$.  Assume a user requests both $X_1$ and $X_2$.  In System (1), the user will not be blocked if they access at least one of the top two drives storing $X_1$, and at least one of the two drives storing $X_2$.  In System (2), the user be will not be blocked and able to decode if they access \textit{any two} drives.  

If there were a fixed independent blocking probability for each drive, then one could show using combinatorics that the System (2)'s blocking probability is lower than System (1)'s.  This type of argument is similar to the distributed storage repair problem discussed in papers such as \cite{DimRamWuSuh:11}.  In this paper, we build upon this idea by firstly having the blocking probability not be independent between drives, but instead a function of the load $ \rho$ and the path multiplicity (cf. (\ref{eq:PbOne})).  This gives a queueing-theory based analysis.  Secondly, we analyze the system under normal operating conditions, with a finer timescale than normal repair problems.  Specifically, in our system, drives can be temporarily and only momentarily blocked and cannot permanently fail.  

In the \ac{SR} system we will explicitly exploit content replication and striping to provide blocking probability savings from replication, without physical replication.  In the \ac{MR} system, we will deal with multiple user types and heterogeneous traffic inflows by adjusting our code to contain systematic components.  Again, this provides the benefits of replication, without physical replication.  

\subsection{General Service Model}
\label{sec:NsysNCS}

We present a general model for the availability and servicing of requested content over time, which can be applied to both single- and multi-resolution analysis.  We model the connecting and service components of hardware that service user read requests: load balancers, servers, I/O buses and storage drives.

A user's read request traverses the following path through  hardware components:   The request arrives at a load balancer and is then  forwarded onto a subset of servers.  Those servers attempt to access connected drives to read out and transfer the requested content back to the user.  Let $ \{S_{y} \}_{y=1}^{n}$ be the set of $n$ \ac{SAN} servers, and $\{D_{y,z}\}_{j=1}^{m^{y}}$ be the $m^{y}$ drives connected to $ S_{y}$.  Define $v$ as the maximum number of drives to which a single server can be connected.

We use the following notation for files and segments.  Let  drives in the \ac{SAN} collectively store a file library $\mathcal{F} = \{ f_{1},\dots,f_{F}\}$, where $f_{i}$ is the $i$th file, and there are $F$ files stored in total.    We order the $T$ segments of file $f_{i}$ by $f_{i}^{(k)} = \{ f_{i,1}^{(k)},\dots,f_{i,T}^{(k)}\}$, where $f_{i,j}^{(k)}$ is the $k$th copy of the $j$th ordered segment of file $i$.  We assume that the \ac{SAN} stores $W$ copies of each file.  

We do not restrict ourselves to files stored on single drives.  In the general case, file segments from the same file copy may be spread across multiple drives.  Striping is an example in which file segments are spread across drives, which can be used to speed up read times  \cite{Far:B00}.  In particular, a server striping file $ f_{i}^{(k)}$  may read sequential segments from the same file in a round-robin fashion among multiple drives.  An example of a common striping standard is the RAID0 standard.     We make the following assumptions about file layout throughout the \ac{SAN}:
\begin{itemize}
\item Define $s$ as the number of drives across which each file is striped; if a file is striped across $s$ drives, we refer to it as an \textit{$s$-striped file}; and
\item The contents of each drive that stores a portion of an $s$-striped file is in the form $f_{i,j}^{(k)}, f_{i,j+s}^{(k)},f_{i,j+2s}^{(k)},\ldots,$ and define these segments as the \textit{$j$th stripe-set for file $f_{i}$}.
\end{itemize}

\begin{figure}[tb]
  \centering
\includegraphics[width = \linewidth]{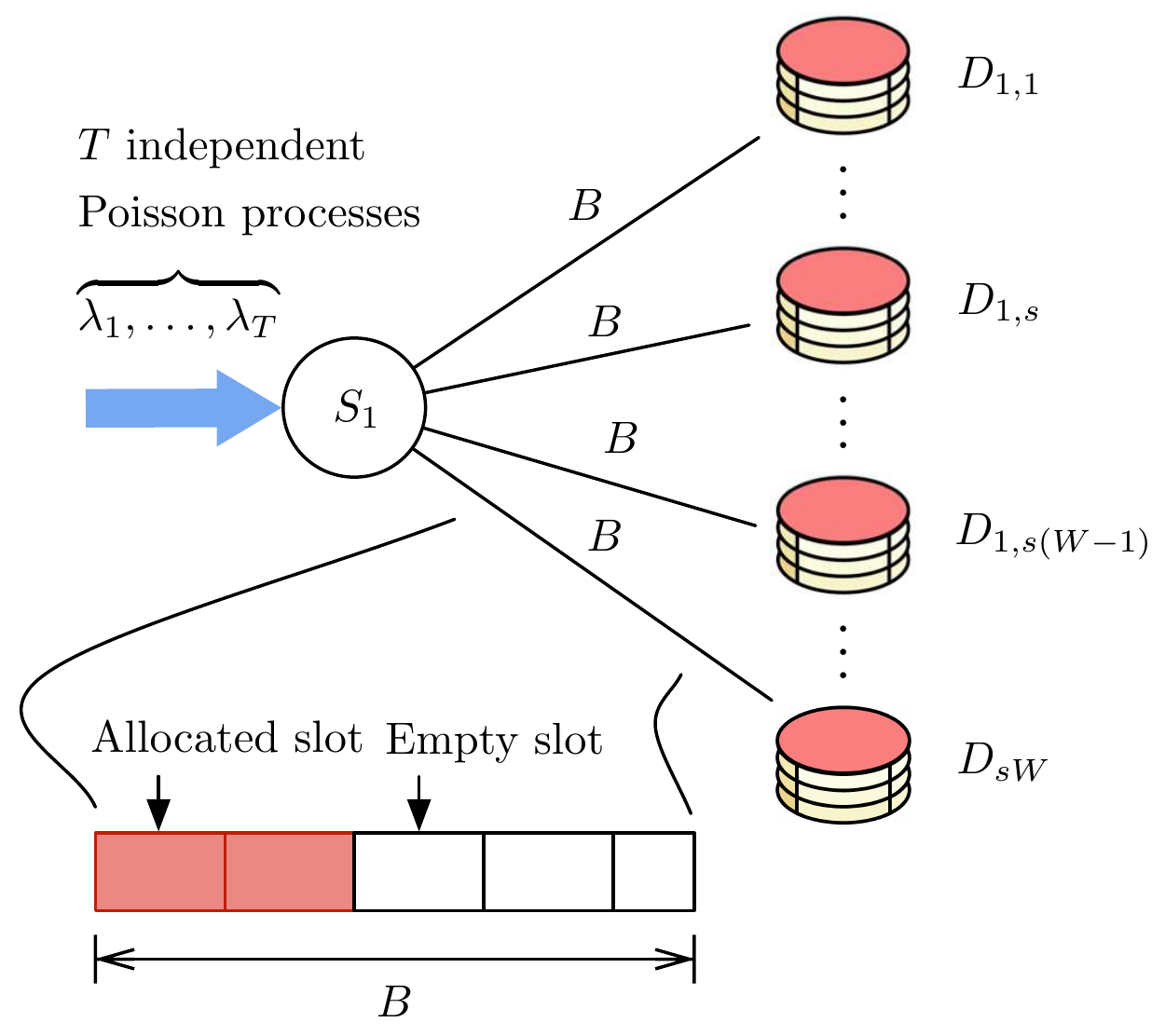} 
\caption{Hardware components in a single-server \ac{SAN} service model.  The server is denoted by $S_{1}$, and drive $(1,z)$ by $D_{1,z}$.  Each drive is connected to a server through an I/O bus with access bandwidth $B$ b/s.  Each segment read request arrives at and is processed by $LB$ prior to being forwarded to some server, and then a set of drives.}
  \label{fig:SANModel}
\end{figure}

Any individual drive can only access a limited number of read requests concurrently; this restriction is particularly pronounced in the case of \ac{HD} video.  Each drive has an I/O bus with access bandwidth $B$ bits/second \cite{Far:B00}, and a download request requires a streaming and fixed bandwidth of $b$ bits/second.  Component connectivity is shown in Fig.~\ref{fig:SANModel}.

When the load balancer receives a request it randomly assigns the request to some server $S_{y}$ with uniform distribution.  (This splits the incoming read request Poisson process and each server sees requests at rate $ \lambda$.)  Server $S_{y}$ then requests the relevant segments from its connected drives $\{D_{y,z}\}_{z=1}^{m^{y}}$.  Since a download request requires a streaming and fixed bandwidth of $b$ bits/second then if the requested file is $s$-striped, each drive I/O bus will require access bandwidth of size $ b/s$.  See Fig.~\ref{fig:SANModel} for an illustration.  In addition, the ratio $sB/b$ is the \textit{number of I/O access bandwidth slots} that each active drive has available to service read arrivals.   Once a particular drive's I/O access bandwidth is allocated, that drive has an average service time of $1/\mu$ seconds to service that segment read  request.  

A drive can only accept a new request from a server if it has sufficient I/O access bandwidth $b/s$ available at the instant the read request arrives.    If a request is accepted by a drive then that drive's controller has determined it can meet the request's read timing guarantees and allocates a bandwidth slot.  Internally, each drive has a disk controller queue for requests and some service distribution governing read request times \cite{ColGru:02,ShrMerWil:98,MerYu:96,Hof:80,Wil:76}.  However, thanks to the internal disk controller's management of request timing guarantees, all accepted request reads begin service immediately from the perspective of the server.\footnote{The full service distribution for modern drives such as SATA drives is dependent on numerous drive model specific parameters including proprietary queue scheduling algorithms, disk mechanism seek times, cylinder switching time, and block segment sizes \cite{MerYu:96}.}   If instead all access bandwidth slots are currently allocated, then that request is rejected or \textit{blocked} by that drive.  If no drives that contain segment $ f_{i,j}$ have available I/O access bandwidth slots, we say that $ f_{i,j}$ is in a \textit{blocked state.}
\begin{definition}
File $f_{i}$ is blocked if there exists at least one segment in $f_{i}$ in a blocked state.  The \textit{blocking probability} for $f_i$ is the steady-state probability that $f_i$ is blocked.  
\end{definition}
See Table \ref{tab:generalModel} for a summary of the general notation used throughout this paper.  

\begin{table}[tb]
  \centering
  \caption{Parameters for general SAN model}
  \label{tab:generalModel}
  \begin{tabular}{|c|l|}
    \hline
    Parameter & Definition \\
    \hline
    $S_y$ & Server $y$ \\
    $m^y$ & Number of drives connected to the $y$th server \\
    $D_{y,j}$ & Drive $j$ connected to server $y$ \\
    $f_{i,j}^{(k)}$ & The $k$th copy of the $j$th segment of file $i$ \\
    $B$ & Maximum I/O access bandwidth for a single drive \\
    $b$ & Bandwidth of each file access request \\
    $s$ & Striping number \\
    $\mu$ & Average drive service rate for segment access requests \\
    \hline
  \end{tabular}
\end{table}

\subsection{Single-resolution Service Model}
\label{sec:nonsyst-serv-model}

In the \ac{SR} service model, we set each file segment to be a file chunk, adhering to video file protocols like \ac{HLS}.  In the \ac{NCS} scheme, coding is performed at the chunk level.  Assume that each file $ f_{i}$ is decomposed into equal-sized chunks, and all files are of the same size.  Typical chunk sizes for video files in protocols such as \ac{HLS} are on the order of a few seconds of playback \cite{HLSDraft:09}, although this is dependent on codec parameters.

Files are not striped among servers, i.e., each file copy $f_{i}^{(k)}$ can be managed by only a
single server, through the results herein can be applied to drives connected to both single or
to multiple servers.  As an example, consider a \ac{SAN} in which $f_{i}^{(1)}$ is striped across three drives.  A server connected to all drives would read segments in the following order: (1) $f_{i,1}^{(1)}$ from $D_{1,1}$; (2) $f_{i,2}^{(1)}$ from $D_{1,2}$; (3) $f_{i,3}^{(1)}$ from $D_{1,3}$; (4) $f_{i,4}^{(1)}$ from $D_{1,1}$, and so on.  

We model user read requests as a set of independent Poisson processes.  In particular, we invoke the Kleinrock Independence Assumption \cite{Kle:B76} and model arriving read requests for each chunk $f_{i,j}$ as a Poisson process with arrival rate $ n \lambda$, which is independent of other chunk read request arrivals.  The Kleinrock Assumption for the independence of incoming chunk read requests requires significant traffic mixing and moderate-to-heavy traffic loads, and so is only appropriate in large enterprise-level \acp{SAN}.

In this paper we use blocking probability, i.e., the steady state probability that there exists at least one chunk in $f_{i}$ that is in a blocked state, as our system performance metric for the \ac{SR} scheme.  

\subsection{Multi-resolution Service Model}
\label{sec:syst-serviceModel}

For the \ac{MR} service model, we present the system model for both the \ac{UCS} and \ac{NCS} \ac{MR} systems. Consider the single server $S_1$ connected to $m^1 = m$ drives.  In the general case, drives store a single  \ac{MR} video that is coded  into $L \in {\cal N}^+$ equal-rate layers, consisting of a base layer $ l_{1}$ and $ L-1$ refinement layers, up to $ l_{L}$.  Each layer has a rate equal to $B_{0}$.  Depending on users' communication bandwidth measurements and resolution preferences, users can request up to $a \in \{ 1, \dots , L \}$ layers, referred to as a \textit{Type $a$} request.  (If a user makes a Type 1 request, they are only requesting the base layer.)  We assume that Type $a$ requests follow an independent Poisson process of rate $ \lambda_a$.  We use an $L$-tuple $(i_1,\dots,i_L)$ to indicate the system state at any given time, where $i_l$ is the number of Type $l$ users currently being serviced in the system.

We assume $L=2$ without loss of generality.  For notational convenience, we set $ i_2=j$ in the paper.  For a given state $ (i,j)$ one of the following four transitions may occur.
\begin{itemize}
\item \textit{Type 1 arrival:} A user requests a base layer $ l_{1}$, modeled as a Poisson process with rate $ \lambda_{1}$; the  process transits to state $(i+1,j)$.
\item \textit{Type 1 departure:} A Type 1 user finishes service and leaves the system with departure rate $ i \mu$; the process transits to state $(i-1,j)$. 
\item \textit{Type 2 arrival:} A user requests  both the base layer $ l_{1}$ and the refinement layer $l_{2}$, modeled as a Poisson process with rate $ \lambda_{2}$; the  process transits to state $(i,j+1)$. 
\item \textit{Type 2 departure:} A Type 2 user finishes service  and leaves the system with departure rate $ j \mu$; the  process transits to state $(i,j-1)$. 
\end{itemize}

The system I/O access bandwidth limits the number of users that can be simultaneously serviced.  In particular, if at the time of arrival of a Type 1 or Type 2 request no relevant drive has sufficient I/O access bandwidth,  then that request is rejected by the system.  If the system can accept neither Type 1 nor Type 2 users, we say that the system is in a \textit{saturated state}.   For a Type 2 request both the base and refinement layers need to be serviced for a block to not occur.  
\begin{definition}
The \textit{saturation probability} $ P_{s}$ is the steady state probability that both Type 1 and Type 2 requests are blocked.
\end{definition}

One could consider a number of different performance metrics similar to saturation probability.  For instance, either Type 1 user saturation or Type 2 user saturation, as opposed to both, could be analyzed.  However, purely Type 1 user saturation is not possible in the \ac{UCS} scheme and Type 2 user saturation does not account for Type 2 users who request content, are rejected, and based on that new information adjust their preferences and re-request Type 1 content.  In contrast to a blocked system, as used in Sec.~\ref{sec:nonsyst-serv-model}, a saturated system represents a more comprehensive form of system unavailability, which we use to avoid the double-counting of Type 2 users who switch down to Type 1 requests.  

\begin{figure}[tb]
\begin{center}
\centering
\includegraphics[width=0.9\linewidth]{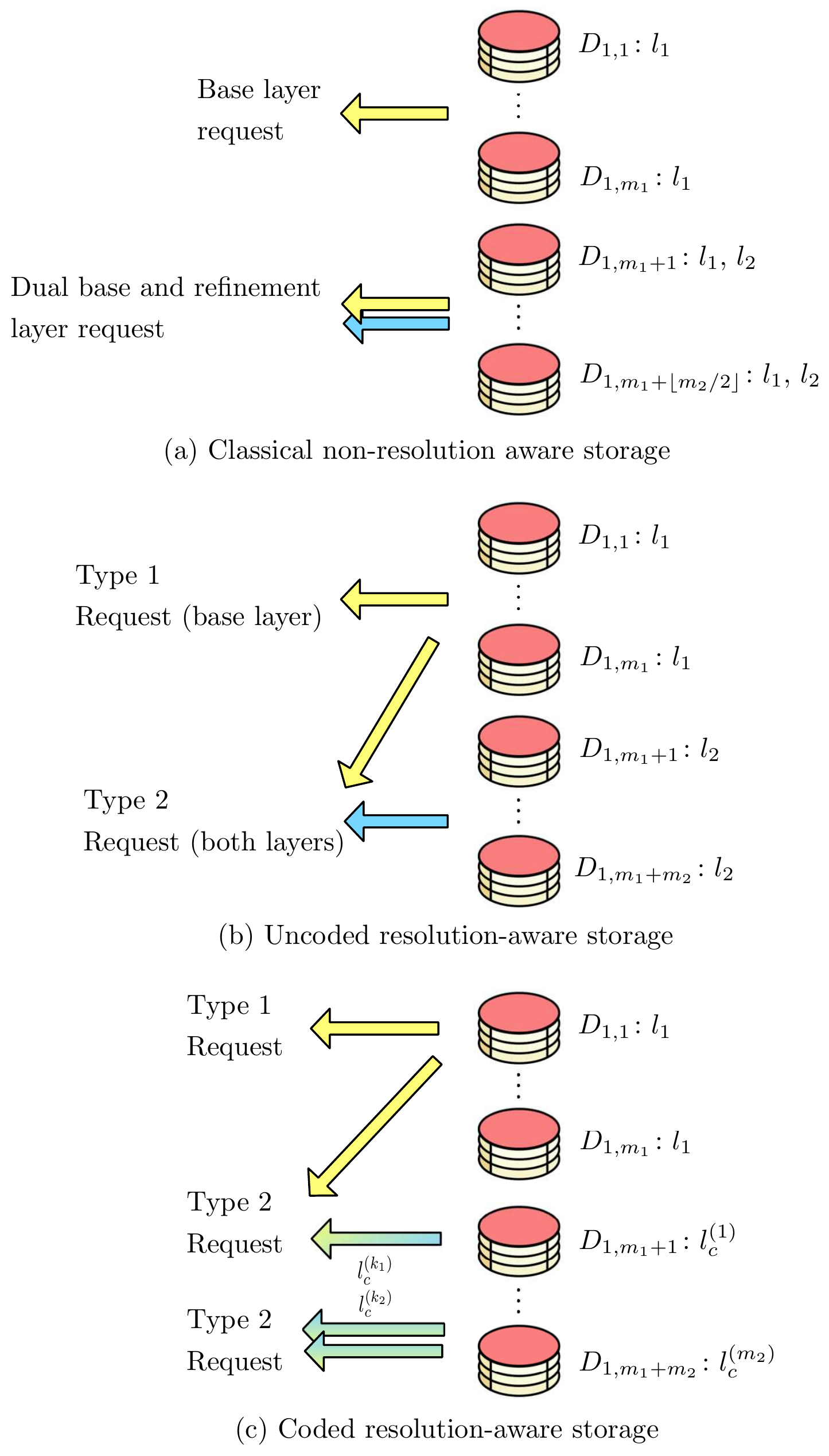}
\end{center}
\caption{In modern storage systems, multi-resolution is usually not exploited at the storage level, where multiple drives store different resolutions of the same video on different drives.  In this non-resolution aware scheme, base and refinement layers for the same resolution are stored on the same drive.  In contrast, in our uncoded resolution-aware storage scheme, Type 2 users, or dual base and refinement layer users, are required to read exactly one  $l_{1}$ copy and one $l_{2}$ copy.  In our coded resolution-aware storage scheme, in addition to \ac{UCS}-style servicing,  Type 2 users can also download two different coded copies from two different drives, both with mixed layers $l_{c}$.}
\label{fig:storage}
\end{figure}

A brief description of the three \ac{MR} schemes follows.  In classical non-resolution aware storage, $m_1 \in \mathbb{N}^{+}$ drives store only base layers $ l_1$ and $\lfloor m_2/2 \rfloor  \in \mathbb{N}^{+}$ drives store both $ l_1$ and $ l_2$.  In the \ac{URS} scheme, $m_{1} \in \mathbb{N}^{+}$ drives store $ l_{1}$ and $ m_{2} \in \mathbb{N}^{+}$ drives store $ l_{2}$, keeping the total storage resources the same as in non-resolution aware storage.  In the \ac{CRS} scheme $ m_{1}$ drives store $ l_{1}$ and $ m_{2}$ drives store random  linear combinations of $ l_{1}$ and $ l_{2}$ as well as the corresponding coefficients.  Let $ l_{c}^{(k)}$ be the $k$th random linear combination,
\begin{align}
  l_c^{(k)} = \sum_{l=1}^L \alpha_l^{(k)} l_l 
\end{align}
where $ \alpha_l^{(k)} \in \mathbb{F}_q$ are coefficients drawn from some finite field of size $q$.  See Fig.~\ref{fig:storage} for illustrations.  We can isolate the storage allocation gains by comparing classical and \ac{URS} schemes.  We can isolate the network coding gains by comparing the \ac{URS} and \ac{CRS} schemes.

In this paper we use saturation probability, i.e., the steady state probability that both Type 1 and Type 2 requests are blocked, as our primary system performance metric for the \ac{MR} schemes.  

\section{Single-resolution Analysis}
\label{sec:analysis}

This section details layout strategies for the two \ac{SR} schemes and presents an analysis of their blocking probability.  We begin by determining the \ac{SAN} blocking probability in a \ac{UCS} scheme as a function of the number of drives and the striping number.  The \ac{NCS} scheme is then described, after which the corresponding blocking probabilities are calculated.    

In the \ac{UCS} scheme, without loss of generality, we set the library $ {\cal F} = \{ f_{i} \}$ to be a single $s$-striped file with $W$ copies of each chunk in the \ac{SAN}.  If no drive contains more than one copy of a single chunk, then $ sW \leq m^{y}, \, \forall y \in \{1,\dots, n \}$.  Assume that all chunks have uniform read arrival rates so $ \lambda = \lambda_{j} \, \forall j \in \{1,\dots,T\}$.  As discussed, the path traversed by each Poisson process arrival is shown in Fig.~\ref{fig:SANModel}, and once a chunk read is accepted by a drive, that drive takes an average time of $ 1/ \mu$ to read the request.

\begin{figure}[tb]
  \centering
\subfigure[An example SAN in which a single server has access to $W$ copies of file $f_{i}$.  Each
connected drive has total  access bandwidth $B$, and each slot takes bandwidth $b/s$.  Each
connected  drive has $\lfloor sB/b \rfloor$ available access bandwidth slots and employs the
\ac{UCS} scheme.   Using \ac{UCS}, the queue mapping for this architecture is shown below in
Fig.~\ref{fige:MGKKQueueMap_UCS}.   For simplicity of illustration, we assume integrality of $T/s$.]{
\includegraphics[width=\linewidth]{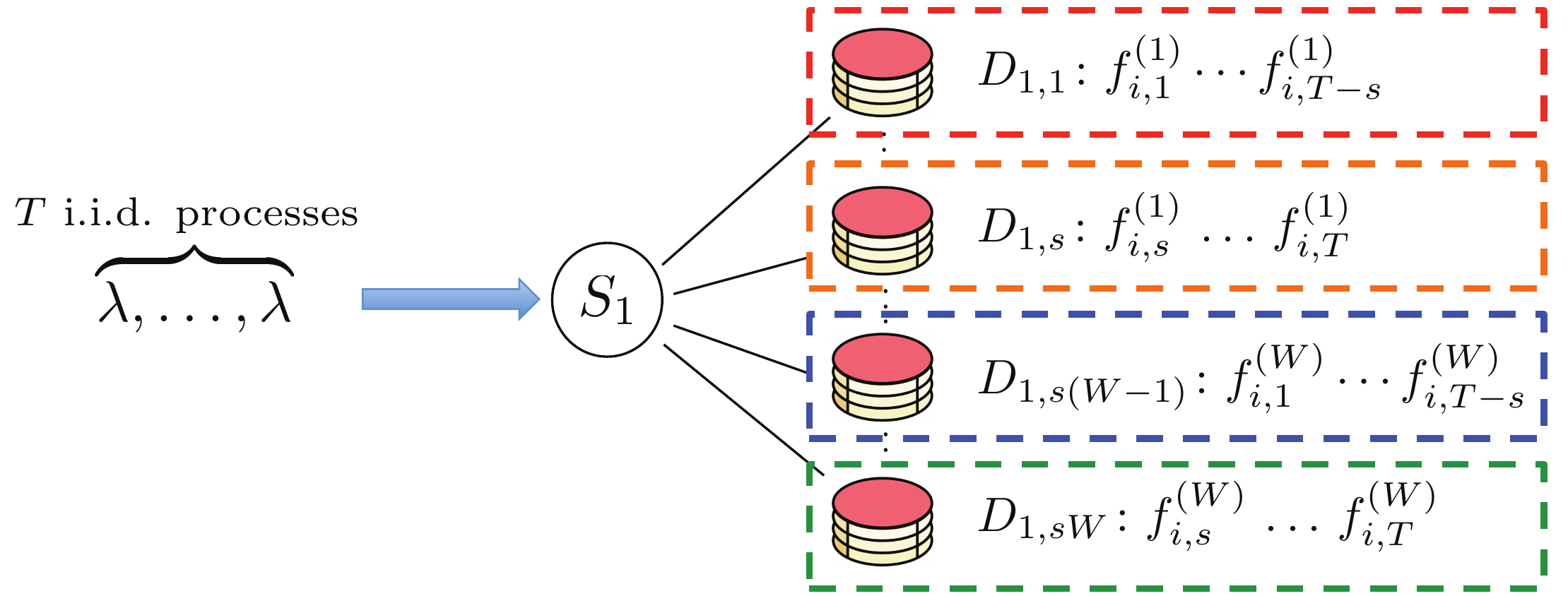}
\label{fig:MGKKQueueHardware_UCS}
}
\\
\subfigure[The equivalent $M/G/K^{U}/K^{U}$ mapping from Fig.~\ref{fig:MGKKQueueHardware_UCS}.  For
simplicity of illustration,  we assume integrality of $T/s$.]{
\includegraphics[width=\linewidth]{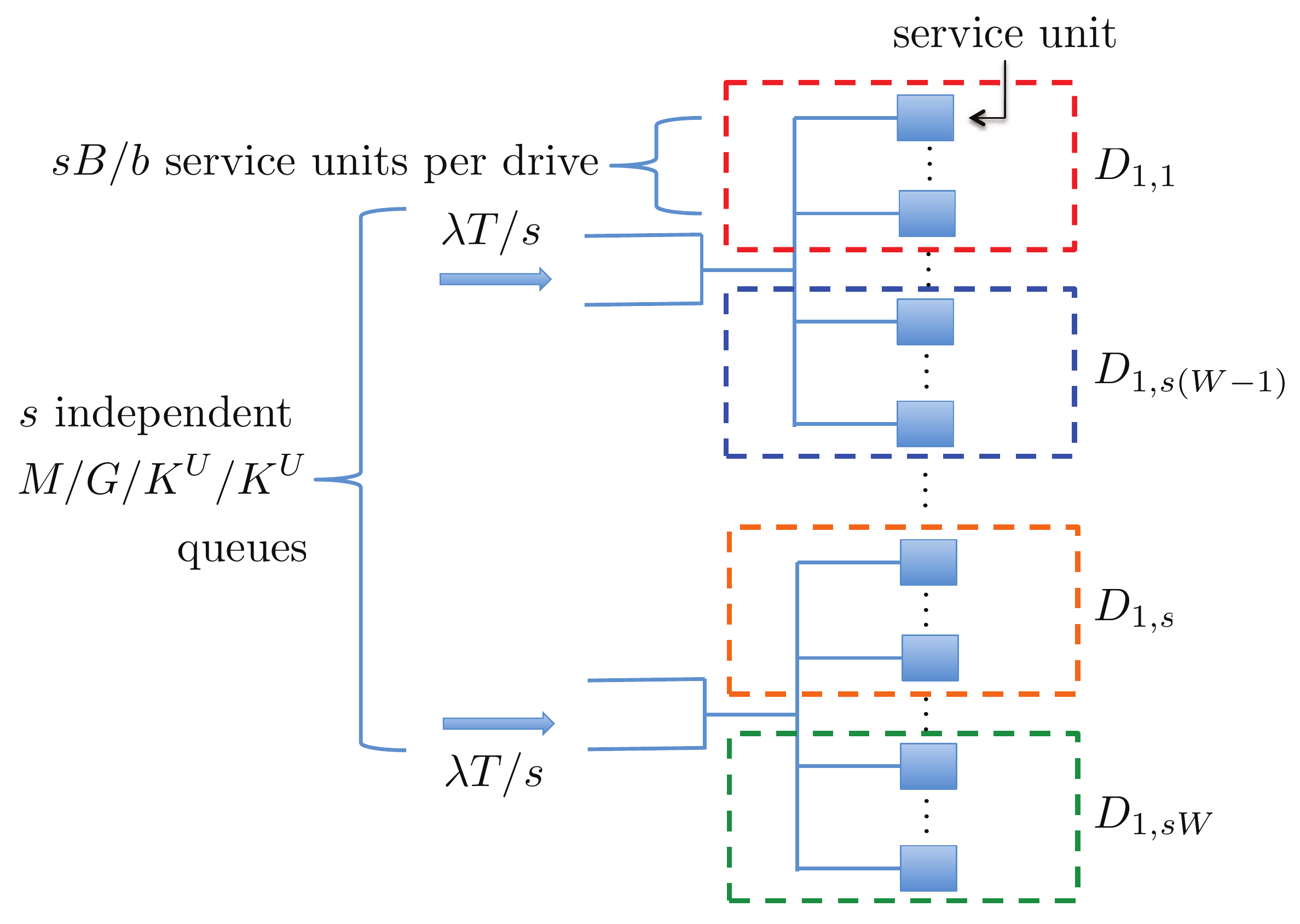}
\label{fige:MGKKQueueMap_UCS}
}

  \caption{An example mapping from a network architecture with a single server to an
    $M/G/K^{U}/K^{U}$ queue in the \ac{UCS} scheme.   In this mapping $s$ independent queues exist.  For  simplicity of illustration, we assume $T \bmod s = 0$.}
  \label{fig:MGKKQueue}
\end{figure}

As per Section \ref{sec:NsysNCS}, file $f_{i}$ is \textit{blocked} if there exists at least one chunk in $f_{i}$ that is in a blocked state.  Chunk $f_{i,j}$ is available if there exists a drive that contains it and has an available access bandwidth slot.   An $s$-striped drive that holds a single stripe set may service requests from either $ \lceil T/s \rceil$ or $\lfloor T/s \rfloor$ different chunks, depending on the length of the stripe set.  We merge all read requests for chunks from the $j$th stripe set of a file copy into a single Poisson process with arrival rate
\begin{align}
\lambda \lfloor T/s \rfloor + \mathbb{I}(j \leq T\bmod s) \, ,
\end{align}
where $j \in \{1, \dots , s \}$ is a drive index containing the $j$th stripe-set, and $ \mathbb{I}(\cdot)$ is the indicator function.  For each file copy, there will be $ T \bmod s $ drives with rate $ \lceil T / s \rceil $ and $ s - T \bmod s $ with rate $ \lfloor T / s \rfloor $.   We map each access bandwidth slot onto a single independent service unit from an $M/G/K^{U}/K^{U}$ queue (see for instance \cite{Kle:B75}) in which each queue has $K^{U}$ service units.  There are $W$ copies of any stripe set on different drives, so our $M/G/K^{U}/K^{U}$ queue has
\begin{align}
  K^{U} = \lfloor sBW/b \rfloor \, 
\end{align}
independent service units for the $j$th stripe set.  This mapping is depicted in Fig.~\ref{fig:MGKKQueue}.  The $M$ denotes that the arrival process is Poisson; $G$ denotes a general service distribution with average rate $\mu $; and $K^{U}$ denotes the total number of service units in the system, as well as the maximum number of active service requests after which incoming requests are discarded \cite{Kle:B75}.  

The blocking probability $P_{b}^{(j)}$ of the $j$th stripe set queue is given by the well-studied \textit{Erlang B} blocking formula,
\begin{align}
  P_{b}^{(j)} & = 
  \begin{cases}
 \frac{ (\rho \lceil T/s \rceil)^{K^{U}} }{ e^{(\rho \lceil T/s \rceil)} \Gamma(1+K^{U},\rho \lceil T/s \rceil) } \, ,\qquad  & j \leq T \bmod s  \,  \nonumber \\ 
\frac{ (\rho \lfloor T/s \rfloor )^{K^{U}} }{ e^{(\rho \lfloor T/s \rfloor)} \Gamma(1+K^{U},\rho \lfloor T/s \rfloor) } \, \qquad & \text{else}  \,  \\ 
  \end{cases}
\end{align}
where $ \rho = \lambda / \mu $ and $\Gamma$ is the upper incomplete Gamma function.  The probability that a chunk is available is equal to $1-P_{b}^{(j)}$.  The probability that $f_{i}$ is blocked $ P_{b}^{U}$, i.e., the probability that not all chunks are available, is then given by
\begin{align}
\label{eq:UCS_Pb}
  P_{b}^{U}  = 1 - &\left( 1- \frac{ (\rho \lceil T/s \rceil)^{K^{U}} }{ e^{\rho \lceil T/s \rceil} \Gamma(1+K^{U},\rho \lceil T/s \rceil) } \right)^{T \bmod s} \nonumber \\
& \times \left( 1- \frac{ (\rho \lfloor T/s \rfloor)^{K^{U}} }{ e^{\rho \lfloor T/s \rfloor} \Gamma(1+K^{U},\rho \lfloor T/s \rfloor) } \right)^{s - T \bmod s} \, .
\end{align}

\subsection{NCS Design}
\label{sec:ncs-design}
\begin{figure}[tb]
  \subfigure[An example of a single server system that has access to only a single copy of file $f_{i}$.  This depiction with $W=1$ is in contrast to Fig.~\ref{fig:MGKKQueue} and is only for visual simplicity.  Chunks are coded using \ac{NCS}, and those in the same highlighted block are composed of coded chunks from the same block window. ]{
    \includegraphics[width=\linewidth]{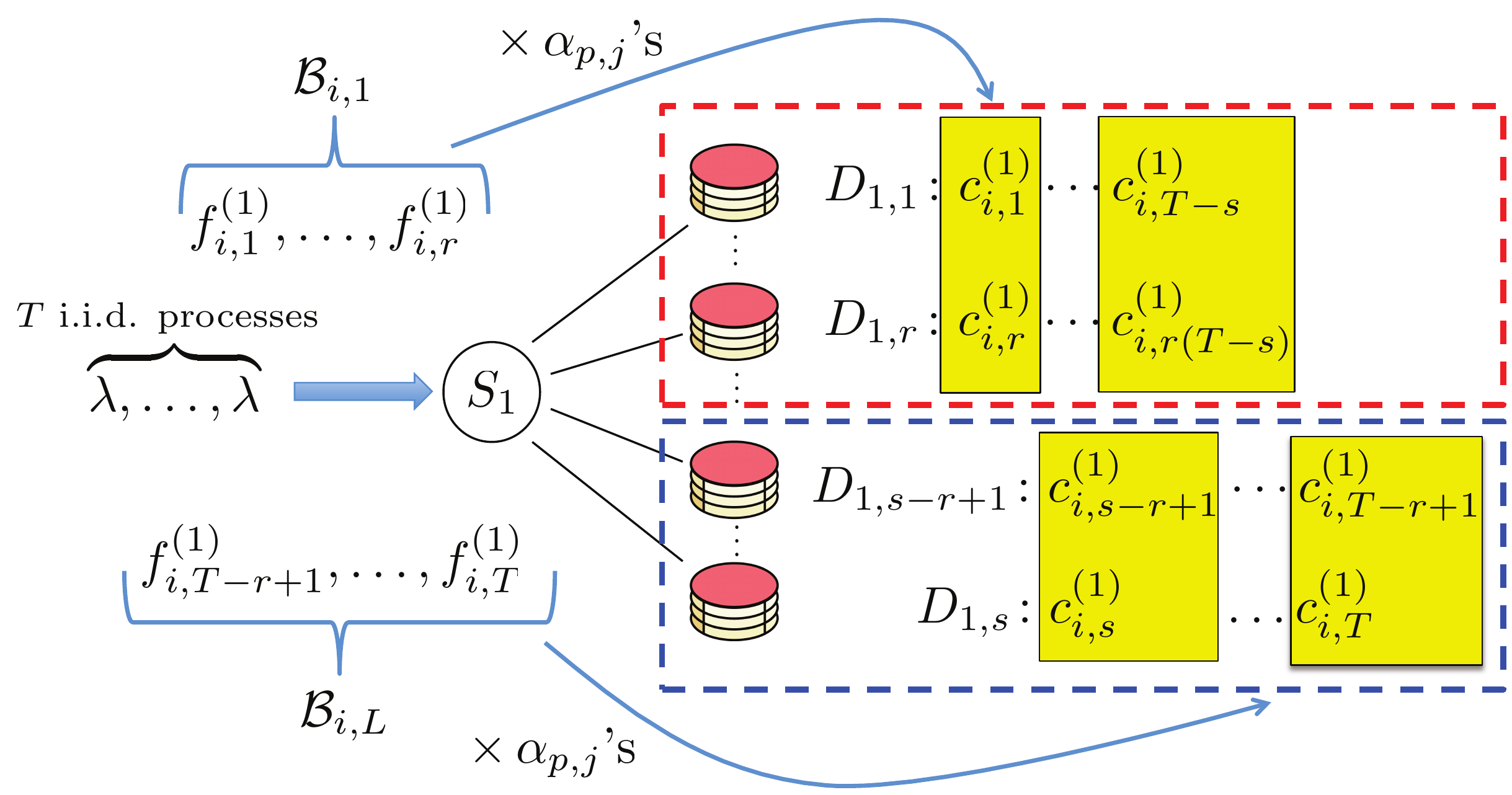}
    \label{fig:MGKKQueueHardware_NCS}
}
\subfigure[A queue mapping from Fig.~\ref{fig:MGKKQueueHardware_NCS}.]{
    \includegraphics[width=\linewidth]{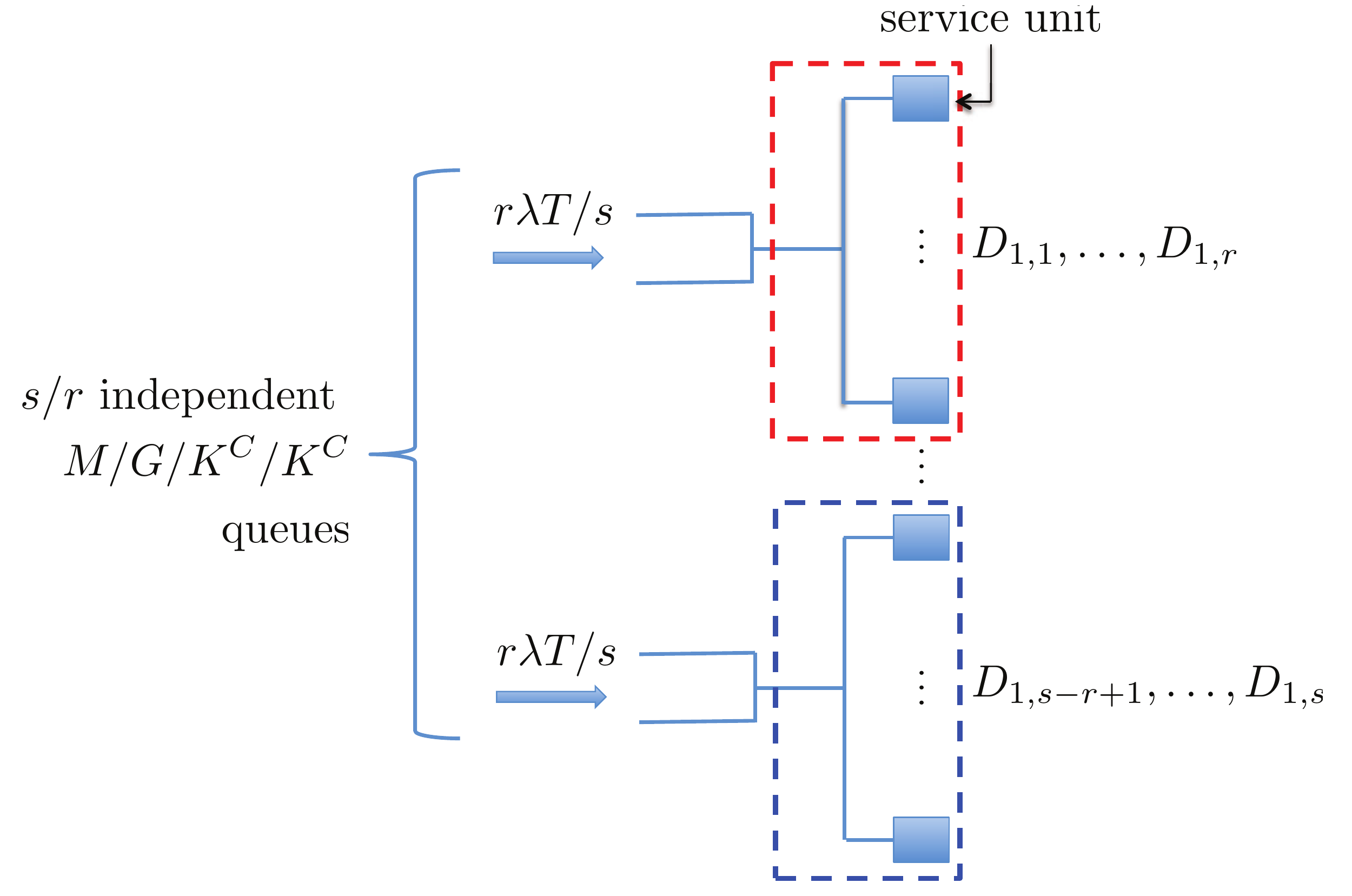}
}
  \caption{An example mapping from a single server hardware architecture with a single copy $W=1$ of file $f_{i}$ to an $M/G/K^{C}/K^{C}$ queue in an \ac{NCS} in blocks of $r$ chunks.  In Fig.~\ref{fig:MGKKQueueHardware_NCS}, file chunks have been coded using \ac{NCS}, and those in the same highlighted block are composed of coded chunks from the same block window as per (\ref{eq:chunkEncode}).  For simplicity of illustration, we assume integrality of $T/s$.}
  \label{fig:MGKKQueue_NCS}
\end{figure}

We now describe our \ac{NCS} scheme and compute the corresponding blocking probability.  \ac{NCS} is equivalent to \ac{UCS} except that we replace each chunk $f_{i,j}^{(k)}$ from the \ac{SAN} with a coded chunk $c_{i,j}^{(k)}$.  To mimic video streaming conditions, we allow a user to receive, decode, and begin playing chunks at the beginning of a file prior to having received the entire file.  Coded chunks are constructed as follows.  We divide each file into equal-sized block windows, or generations, each containing $r$ chunks.  Owing to striping, we constrain $r \leq s$ and $s/r \in \mathbb{N}^{+}$.  (There will be no performance gain from network coding if there is coding across chunks on the same drive, and coding across chunks on the same drive must exist if $r > s$.)  Let ${\cal B}_{i,l}$ be the $l$th block window/generation, where ${\cal B}_{i,l}$ is a subset of file $ f_{i}$'s chunk indices and $ {\cal B}_{i,l}$ is disjoint from all other block windows.  See Fig.~\ref{fig:MGKKQueueHardware_NCS} for an illustration.  

Coded chunk  $c_{i,j}^{(k)} \, , \, \, \, j \in {\cal B}_{i,l}$, is a linear combination of all uncoded chunks in the same block window that contains $f_{i,j}$,
\begin{align}
  c_{i,j}^{(k)} = \sum_{p \in B_{i,l} } \alpha_{p,j}^{(k)} f_{i,p}^{(k)} \, 
\label{eq:chunkEncode}
\end{align}
where $\alpha_{p,j}^{(k)}$ is a column vector of coding coefficients drawn from a finite field $\mathbb{F}_{q}$  of size $q$ \cite{KoeMed:03}, and where we treat $ f_{i,p}^{(k)}$ as a row vector of elements from $ \mathbb{F}_{q}$.  We assign coding coefficients that compose each $ \alpha_{p,j}^{(k)}$ with uniform distribution from $ \mathbb{F}_{q}$, mirroring \ac{RLNC} \cite{HoMedKoeEffShiKar:06}, in which the random coefficients are continuously cycled.  In this scheme, coded chunk $c_{i,j}$ now provides  the user with partial information on all chunks in its block window.  Note that coefficients are randomly chosen across both the chunk index $j$ as well as the copy $k$.  Similarly to \cite{AceDebMedKoe:05}, when a read request arrives for a coded chunk, the relevant drive transmits both $c_{i,j}^{(k)}$ as well as the corresponding coefficients $\{ \alpha_{p,j}^{(k)} \}$.  
In such systems complexity is low because inverting small matrices is not energy intensive or particularly time consuming \cite{AceDebMedKoe:05}.  

In the \ac{NCS} scheme, the blocking probability is calculated as follows.  Similar to \ac{UCS}, we merge the independent Poisson arrival processes for uncoded chunks $\{ f_{i,j} \colon j \in B_{i,l} \}$ into a Poisson process with arrival rate either $ r \lambda \lceil T/s \rceil$ or $ r \lambda \lfloor T/s \rfloor$, depending on the stripe-set length.  This process can be interpreted as requests for any coded chunk that has an innovative degree of freedom in the $l$th block window.  See Fig.~\ref{fig:MGKKQueue_NCS} for an example mapping from a hardware architecture to a queue in which $W=1$.  Generalizing such an architecture, the request rates for an innovative chunk in the $l$th block window are again mapped to an equivalent $M/G/K^{C}/K^{C}$ queue with parameter
\begin{align}
  K^{C} = r K^{U} \, ,
\end{align}
and so the blocking probability $P_{b}^{(jC)}$ for each coded stripe set $ M/G/K^{C}/K^{C}$ queue is given by 
\begin{align}
  P_{b}^{(jC)} & = 
  \begin{cases}
    \frac{ (r \rho \lceil T/s \rceil )^{K^{C}}}{ e^{r \rho  \lceil T/s \rceil} \Gamma(1+K^{C},r\rho \lceil T/s \rceil) } \, , \qquad  & j \leq \left(T \bmod s \right) /r  \,  \nonumber \\  
    \frac{ (r \rho \lfloor T/s \rfloor )^{K^{C}}}{ e^{r \rho  \lfloor T/s \rfloor} \Gamma(1+K^{C},r\rho \lfloor T/s \rfloor) } \, , \qquad  & \text{else}  \, .
  \end{cases}
\end{align}
The constraint $s/r \in \mathbb{N}^{+}$ ensures that the queueing model has  $s/r$ independent queueing systems and that no intra-drive coding exists, as with the \ac{UCS} scheme.  The \ac{NCS} blocking probability $P_{b}^{C}$ is given by
\begin{align}
\label{eq:NCS_Pb}
    P_{b}^{C}  & = 1 - \left( 1- \frac{ (r \rho \lceil T/s \rceil )^{K^{C}}}{ e^{r \rho  \lceil T/s \rceil} \Gamma(1+K^{C},r\rho \lceil T/s \rceil) } \right)^{\frac{ T\bmod s}{r}} \nonumber \\
& \times \left( 
1 - \frac{ (r \rho \lfloor T/s \rfloor )^{K^{C}}}{ e^{r \rho  \lfloor T/s \rfloor} \Gamma(1+K^{C},r\rho \lfloor T/s \rfloor) }
\right)^{\frac{ s }{ r }  - \frac{ T \bmod s }{  r } }\, .
\end{align}

\subsection{UCS and NCS Comparison}
\label{sec:UCSComparison}

\begin{figure}[htb]
   \centering
\subfigure[The effect of \ac{NCS} on duplication requirements in a low stripe-rate system with low load.  In this setup the stripe-rate is set to $ s=2$, the number of chunks is $T=150$, $B/b= 2$ and $ \rho = 0.2$.]{
\psfrag{Pb}{$P_{b}$}
\psfrag{W}{$W$}
\psfrag{r=2}{$ r=2$}
\includegraphics[width=0.65\linewidth]{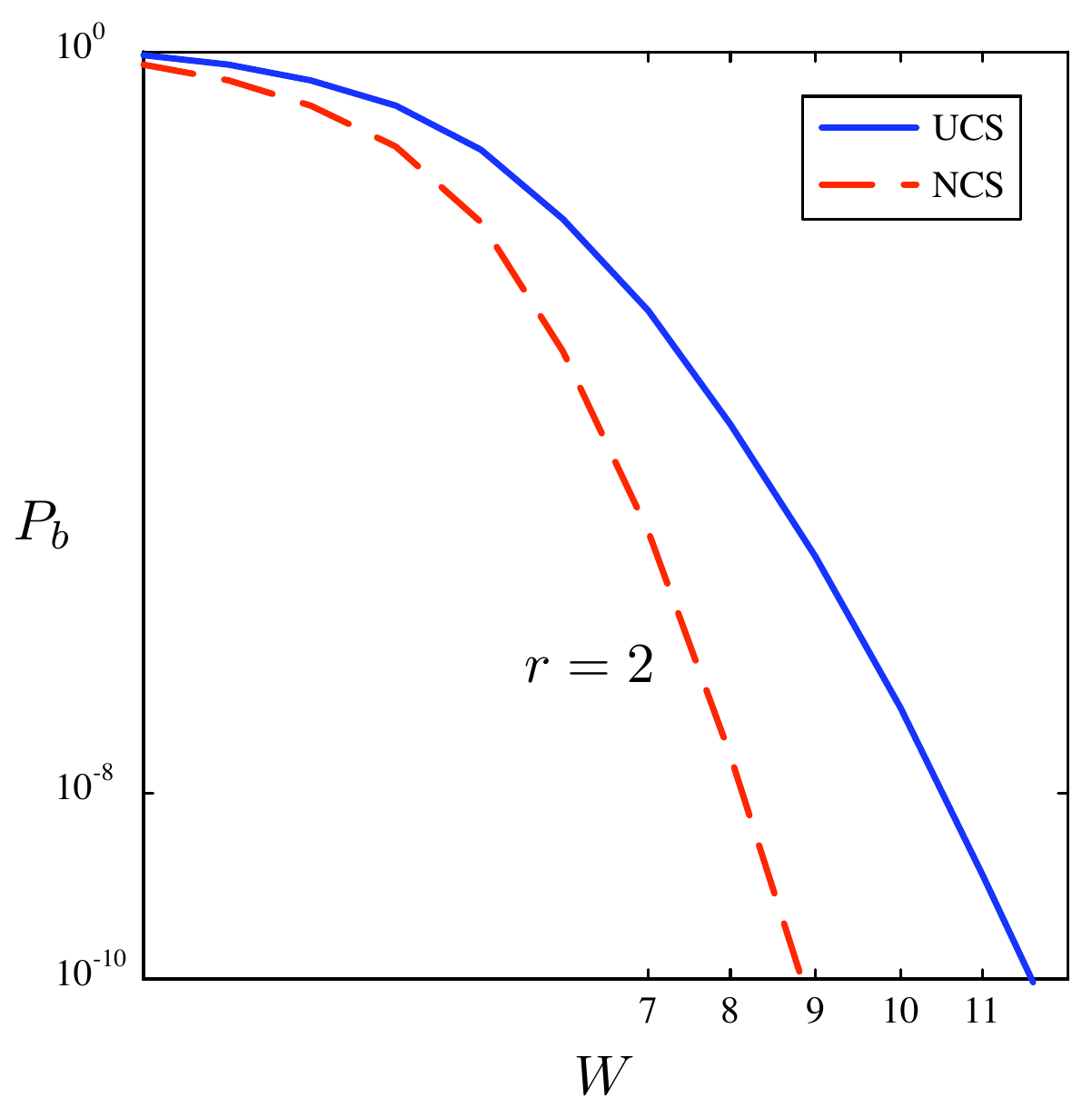}
} \\
\subfigure[A medium stripe-rate in a system with heavy load.  In this setup the
stripe-rate is set to $ s=4$, the number of chunks is $T=150$, $B/b= 2$ and $ \rho = 0.9$.  The
arrow notation on \ac{NCS} curves denotes the direction that curves are labeled when the label is
read from left-to-right. ]{
\psfrag{Pb}{$P_{b}$}
\psfrag{W}{$W$}
\psfrag{r=2,4}{$ r=2,4$}
\includegraphics[width=0.65\linewidth]{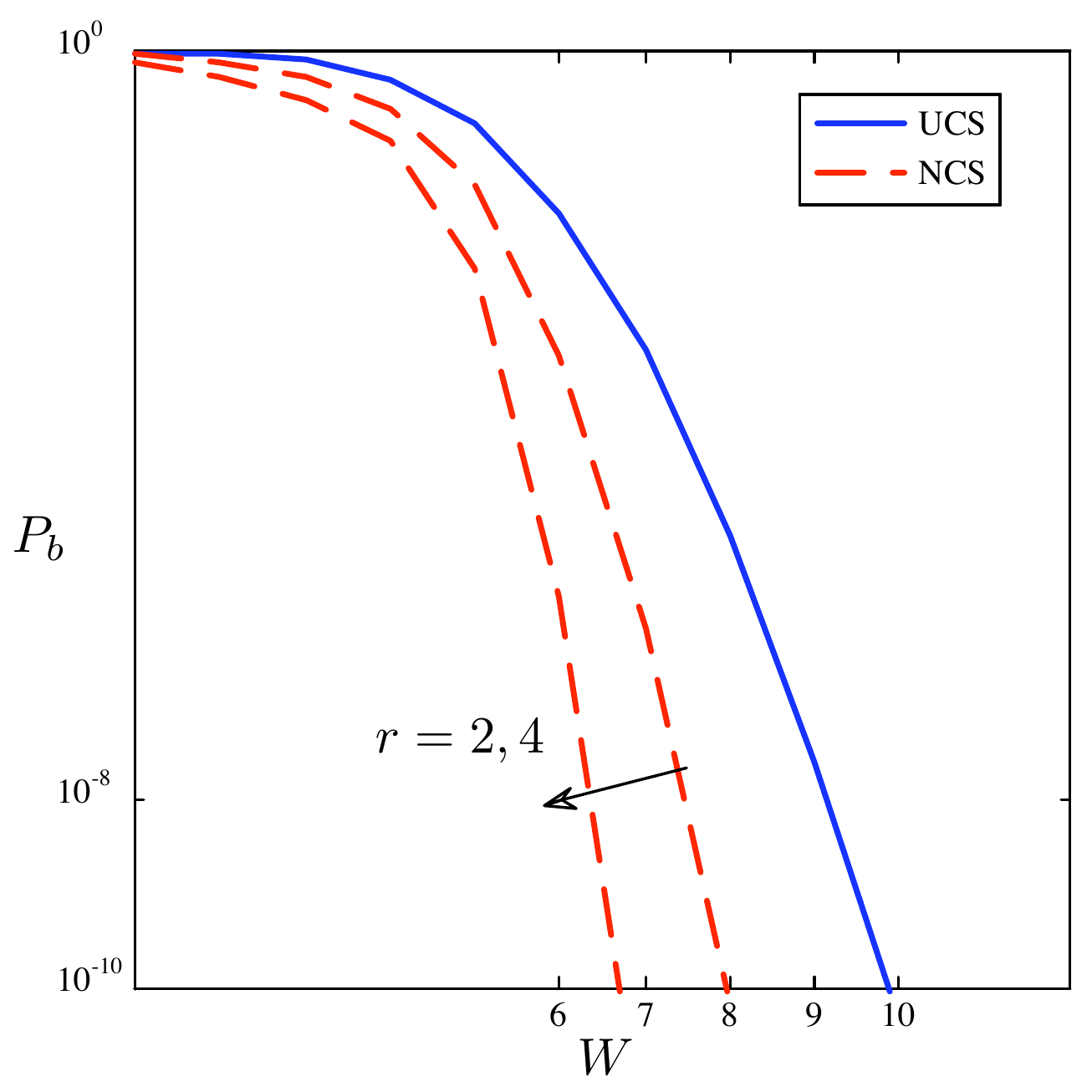}
} 
\caption{The effect of \ac{NCS} on duplication requirements as a function of blocking probability
  under various stripe-rates and system loads.}
\label{fig:blockingProbabilities}
\end{figure}

\begin{figure}[tb]
\centering
\includegraphics[width=0.65\linewidth]{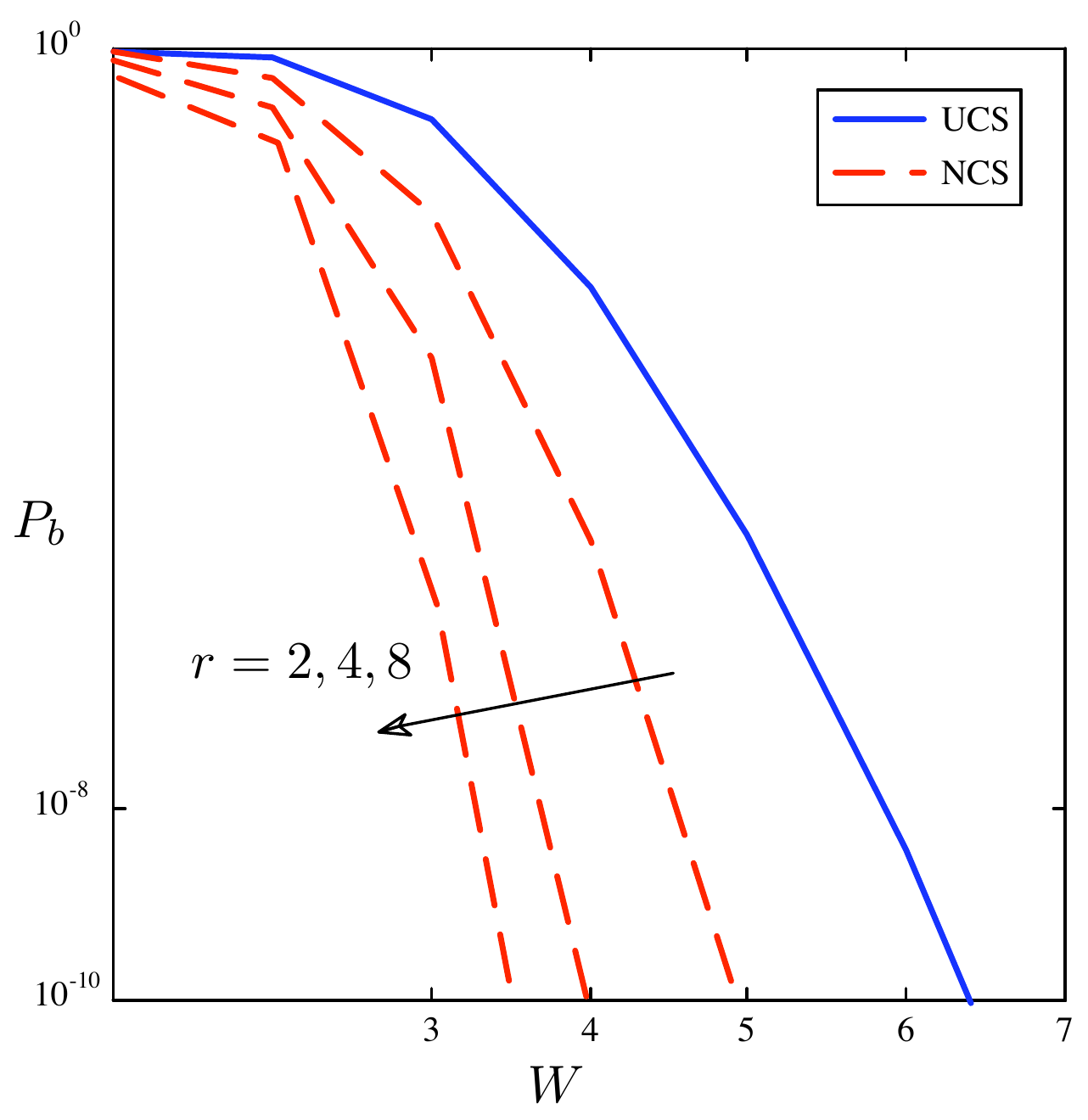}
\caption{The effect of \ac{NCS} on duplication requirements as a function of blocking probability
  under a high stripe-rate.  In this setup the stripe-rate is set to $ s=8$, the number of chunks is $T=150$, $B/b= 1$ and $ \rho = 0.9$.}
\label{fig:blockingProbabilities_highStripe}
\end{figure}

We now compare the blocking probabilities of the \ac{SR} \ac{NCS} and \ac{UCS}.  Figs.~\ref{fig:blockingProbabilities} and \ref{fig:blockingProbabilities_highStripe} plot (\ref{eq:UCS_Pb}) and (\ref{eq:NCS_Pb}) as a function of  $W$ for three different stripe-rates $s=2,4,8$, which we refer to as low, medium, and high stripe-rates, respectively. We set the number of chunks to $T=150$ to approximate a short movie trailer if chunks are divided up using a protocol such as \ac{HLS} \cite{HLSDraft:09}.  Finally, the number of videos that each drive can concurrently service $B/b$ is set to 2.

As the stripe-rate increases, the benefit of the \ac{NCS} scheme over \ac{UCS} becomes more apparent. In particular, assuming a target \ac{QOS} of $P_{b} = 10^{-8}$, the low stripe-rate scenario requires 20\% fewer file copies.  In contrast, the high stripe-rate scenario requires approximately 50\% fewer copies.

\section{Multi-resolution Analysis}
\label{sec:systematicNCS}

This section presents a Markov process system model for the three \ac{MR} schemes.  Numerical results are then used to compare their saturation probabilities.

\begin{figure*}[tb]
  \centering
\includegraphics[width=0.9\linewidth]{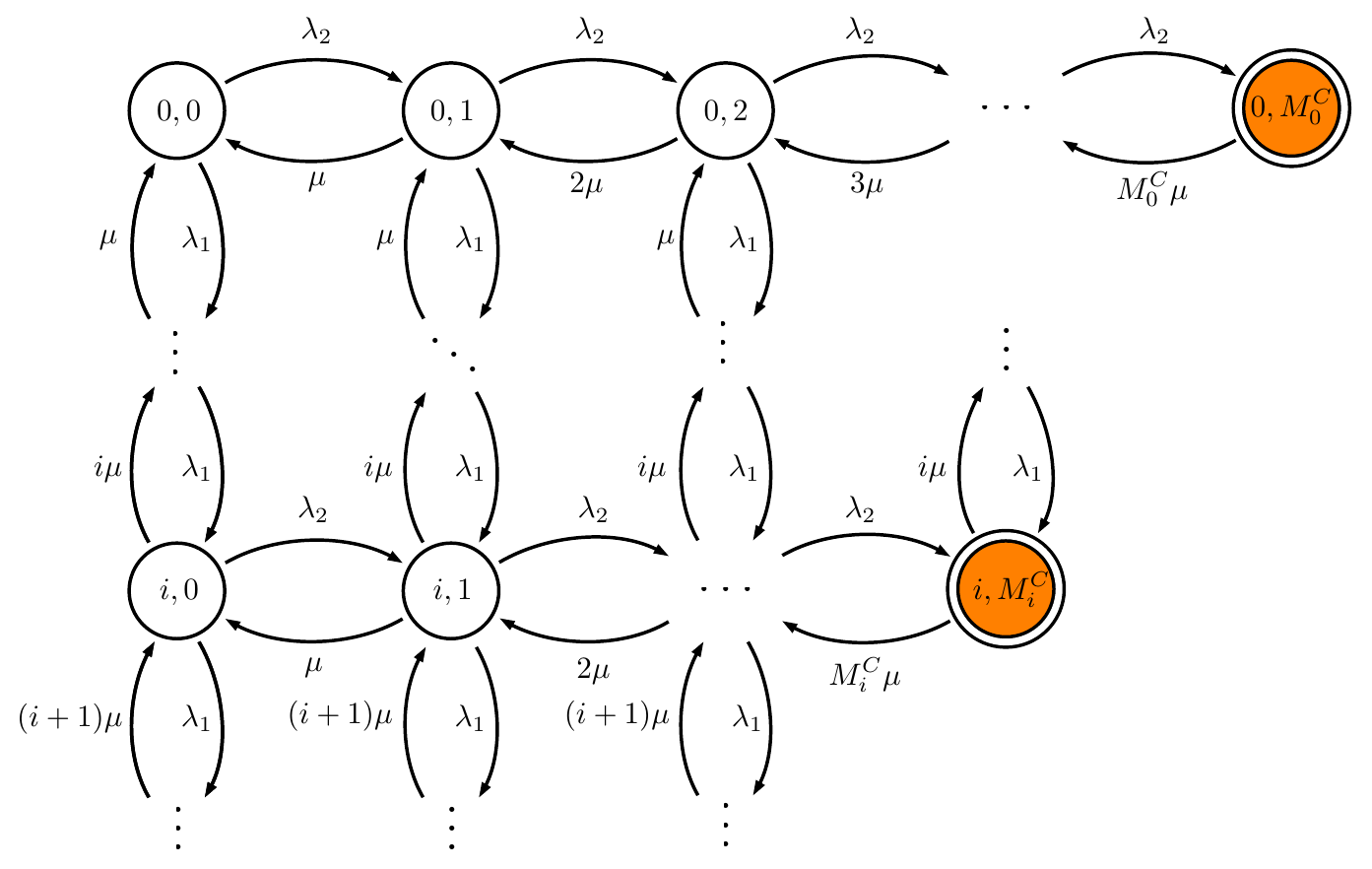}
\caption{An example two-dimensional Markov process modeling the \ac{CRS} scheme, as described in Section \ref{sec:model}.  State $ (i,j)$ denotes $i$ Type 1 and $j$ Type 2 users in the system.  The set of saturation states $ {\cal B}$ are shaded.}
\label{fig:MCEta5}
\end{figure*}

As a reminder, for this analysis we use the notation presented in Section \ref{sec:syst-serviceModel}.  We give an example of a general \ac{CRS} Markov process in Fig.~\ref{fig:MCEta5}, in which $i$ is the number of Type 1 users  and $j$ is the number of Type 2 users in the system.  Define the set of saturation states as $ {\cal B}$.  For the set of shaded saturation states  ${\cal B}$, given $i$ Type 1 users in the system, let $ M_{i}^{U}$ and $ M_i^{C}$ be the maximum number of Type 2 users  that can be concurrently serviced in the \ac{UCS} and \ac{NCS} schemes, respectively.  To capture the essence of the differences among the classical, \ac{URS}, and \ac{CRS} schemes, we assume perfect scheduling and focus instead on the boundaries of each Markov process.  

\subsection{System Constraints \& Select Analysis}
\label{sec:systemConstraintsClassicAnal}

This subsection outlines the system constraints for the three \ac{MR} schemes and presents a simple analysis technique for the classic non-resolution aware scheme.  First, consider the classical non-resolution aware scheme.  The constraints for base and refinement layers are of the form   $iB_0 \leq m_1B$, so
\begin{align}
  i \leq m_1 \lfloor B/B_0 \rfloor
\end{align}
and 
\begin{align}
  j \leq m_2 \lfloor B/B_0 \rfloor \, .
\end{align}
Similar to Section \ref{sec:analysis}, the maximum number of users is equivalent to the number of service units in an $M/G/K/K$ queue.  In addition, since the Type 2 files are twice as large, we halve their service rate.  The same technique that was used to generate (\ref{eq:UCS_Pb}) gives the \ac{MS}  saturation probability $ P_s^R$ as
\begin{align}
\label{eq:PsR}
  P_s^R = & \frac{ (\lambda_1 / \mu )^{m_1 \lfloor B/B_0 \rfloor}/ (m_1 \lfloor B/B_0 \rfloor)! }{\sum_{i=0}^{m_1 \lfloor B/B_0 \rfloor} (\lambda_1/\mu)^i/i!} \nonumber \\ 
 & \times  \frac{ (2 \lambda_2 / \mu )^{m_2 \lfloor B/B_0 \rfloor}/ (m_2 \lfloor B/B_0 \rfloor)! }{\sum_{i=0}^{m_2 \lfloor B/B_0 \rfloor} (2 \lambda_2/\mu)^i/i!} \, .
\end{align}

Second, consider the \ac{URS} scheme.  The drive I/O access bandwidth constraints sets the Markov process boundaries and defines $ M_{i}^{U}$.  In particular, we have the following constraints:
    \begin{align}
\label{eq:UCSConstraints}
 L_1 \text{ storage, } &  (i+j)B_0\leq m_{1}B \,   \nonumber \\ 
L_2  \text{ storage, } & jB_0\leq m_{2}B \,  \\ 
\text{ Total storage, }& (i+2j)B_0\leq(m_1+m_2)B \nonumber \,  
\end{align}
which implies
\begin{align}
  M_{i}^{U} = \min 
\begin{cases} 
 m_{2} \lfloor \frac{ B }{ B_{0} }  \rfloor \, , \\
 m_{1}  \lfloor \frac{ B }{ B_{0} } \rfloor - i \, , \\
 \frac{ m_{1}+m_{2} }{ 2 }  \lfloor    \frac{ B }{ B_{0} } \rfloor  - \lceil i/2 \rceil \, .
\end{cases}
\end{align}
Note that in this scheme if $ m_{2} > m_{1}$ then not all I/O access bandwidth will be simultaneously usable.  Specifically, not all Type 2 access bandwidth slots will be usable because each refinement layer requires an accompanying base layer, and there are  more refinement than base layers stored.

\begin{figure}[tb]
  \centering
  \subfigure[A comparison of saturation states $ {\cal B}$ for the \ac{URS} and \ac{CRS} Markov processes.]{
  \includegraphics[width=0.8 \linewidth]{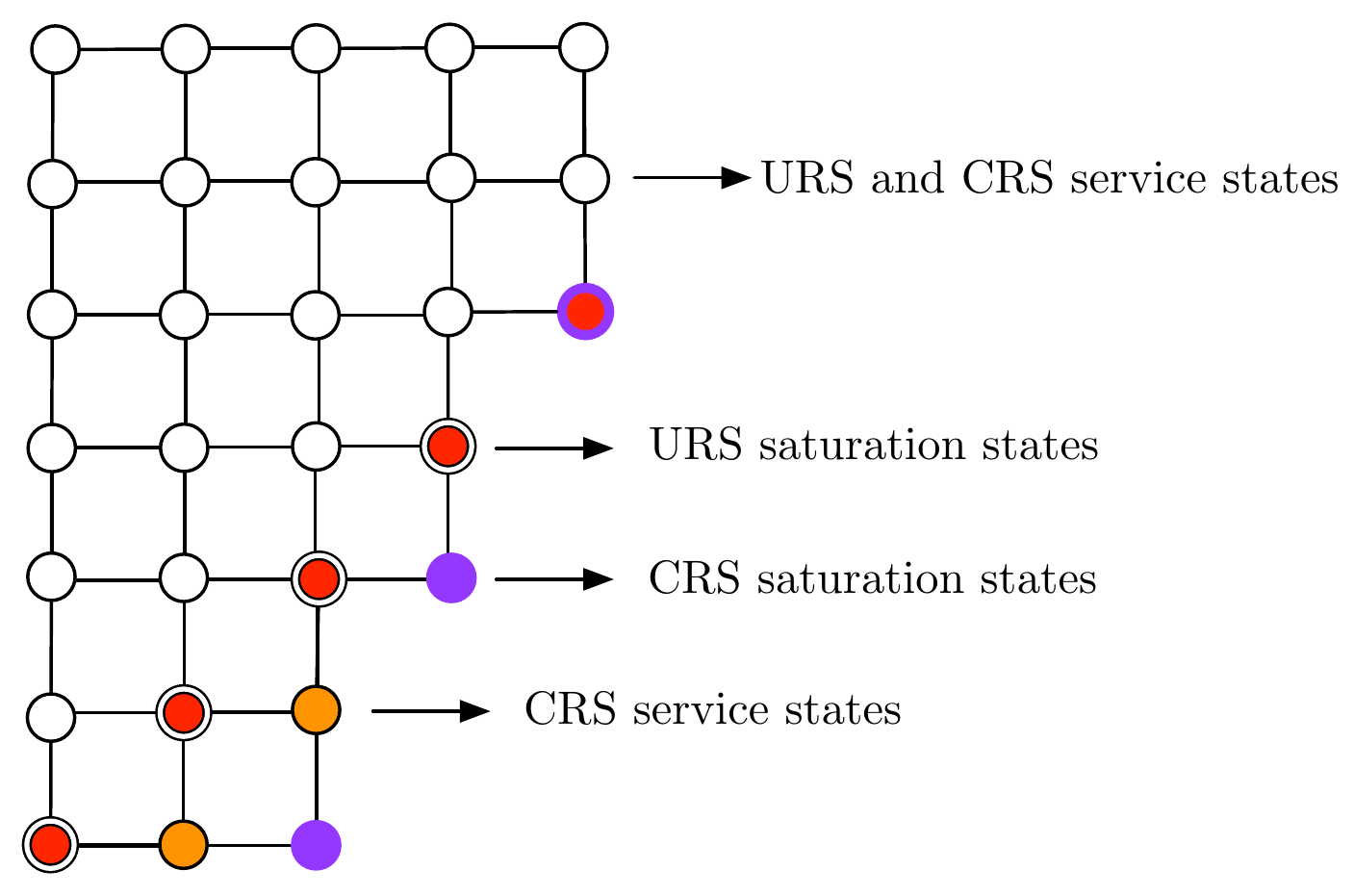}
}
  \subfigure[A comparison of the process boundaries for the \ac{URS} and \ac{CRS} Markov processes.]{
    \label{fig:MPBoundariesSubFig}
  \includegraphics[width=0.8 \linewidth]{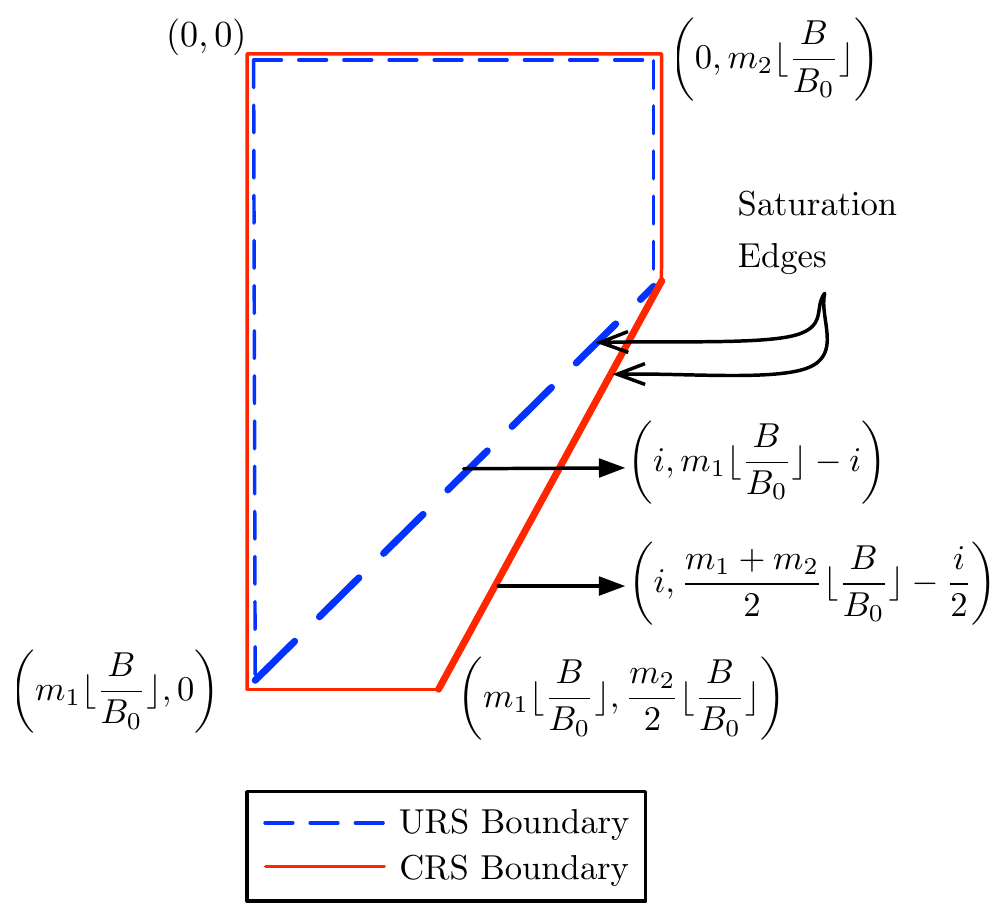}
}
  \caption{A comparison of the lattice structures for the \ac{URS} and \ac{CRS} scheme Markov processes.  The bottom right-hand diagonal edges indicate where the system is saturated.}
  \label{fig:MPExampleAndBoundaries}
\end{figure}

Third, consider the \ac{CRS} scheme.    The drive I/O access bandwidth constraints again sets the Markov process boundaries and defines $ M_{i}^{C}$.  Similar to (\ref{eq:UCSConstraints}), we have the following constraints:
    \begin{align}
\label{eq:NCSConstraints}
 L_1 \text{ storage, } &  iB_0\leq m_{1}B \,   \nonumber \\ 
L_2  \text{ storage, } & jB_0\leq m_{2}B \,  \\ 
\text{ Total storage, }& (i+2j)B_0\leq(m_1+m_2)B \, \nonumber 
\end{align}
which implies
\begin{align}
  M_{i}^{C} = \min 
\begin{cases} 
 m_{2} \lfloor \frac{B }{ B_{0} }  \rfloor \, , \\
 m_{1}  \lfloor \frac{B }{ B_{0} } \rfloor  \, , \\
\frac{ m_{1}+m_{2} }{ 2 }  \lfloor \frac{ B }{ B_{0} } \rfloor  - \lceil i/2 \rceil \, .
\end{cases}
\end{align}
The number of states in the \ac{CRS} process is larger than the \ac{URS} process, and  $ M_{i}^{C} \geq M_{i}^{U}$, as shown in Fig.~\ref{fig:MPExampleAndBoundaries}.  This quantifies the additional degrees of freedom provided by \ac{CRS} from more options to service Type 2 requests.  

The saturation probability of the \ac{URS} and \ac{CRS} Markov processes requires computing partial steady-state distributions for states in $ {\cal B}$.  Consider the simpler \ac{URS} scheme.  The trapezoidal shape of process boundaries yields $ {\cal O}(K_{1}M_{0})$ equations to solve this distribution, where $ K_{1}$ is the globally maximum numbers of Type 1 users that the system can accept.  The trapezoidal boundaries are a variation of a related Markov process with rectangular boundaries, i.e., an identical process in which the triangle wedge, as depicted in Fig.~\ref{fig:MPBoundariesSubFig}, has not been removed.  In such a system, saturation probabilities could be computed using  results very similar to those found in \cite{FaKinMit:82}.  However, the systematic storage used by the \ac{CRS} scheme restricts our ability to use such techniques.  Instead, we explore the opportunity for gains from the \ac{CRS} scheme using numerical results.  

\subsection{Numerical Results}
\label{sec:results}

In this section, we consider Monte Carlo based numerical simulations that compare saturation probability $P_{s}$ of the \ac{URS} and \ac{CRS} schemes.  In particular, we consider the effect of the server load, and the type request ratio given an optimal allocation strategy.

\subsubsection{Server Load}

\begin{figure}[tb]
\begin{center}
\centering
\psfrag{Pb}{$P_{s}$}
\psfrag{rho}{$\lambda / \mu $}
\includegraphics[width=0.85\linewidth]{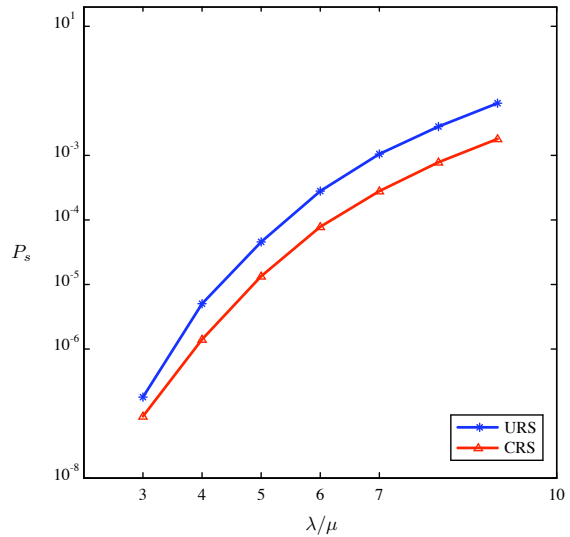}
\end{center}
\caption{The effect of the server load on the saturation probability $P_{s}$.  In the setup, $m = 12, \, m_{1} = 8, \, \lambda_{1} = \lambda_{2}, \, \lambda = \lambda_1 + \lambda_2, \, B/B_{0} = 2$.}
\label{fig:Ps_rho}
\end{figure}

In this subsection we assume symmetric arrival processes so that $ \lambda = \lambda_{1} + \lambda_{2}$, where $ \lambda_1 = \lambda_2$, and we compare $P_{s}$ for the \ac{URS} and \ac{CRS} schemes as a function of $\lambda / \mu$.  This provides insight into how the server load affects  saturation probability of the two schemes.  As a case study, consider a system in which each drive can concurrently stream two layers, i.e., $B / B_{0} = 2$, and in which the single server is connected to a standard $m=12$ drives, where $m_{1} = 8$ drives  contain $L_{1}$.  (Under the \ac{URS} scheme, Type 2 requests have double the bandwidth requirements of Type 1 requests, so we set $ m_{1} = 2m_{2}$.)  Fig.~\ref{fig:Ps_rho} plots the results.  

There is only a modest and relatively constant $P_{s}$ gain for  \ac{CRS} over the \ac{URS} scheme.  We are interested in avoiding totally blocked system states, as per our $ P_{s}$ definition.  Hence, for the remainder of the paper we consider a fully-loaded system in a near-constant saturation state, i.e., with a very high saturation probability.  As such, we set $ \lambda / \mu = 6$ for the remainder of this paper.

\subsubsection{Type Request Ratio}

\begin{figure}[tb]
  \centering
  \includegraphics[width = \linewidth]{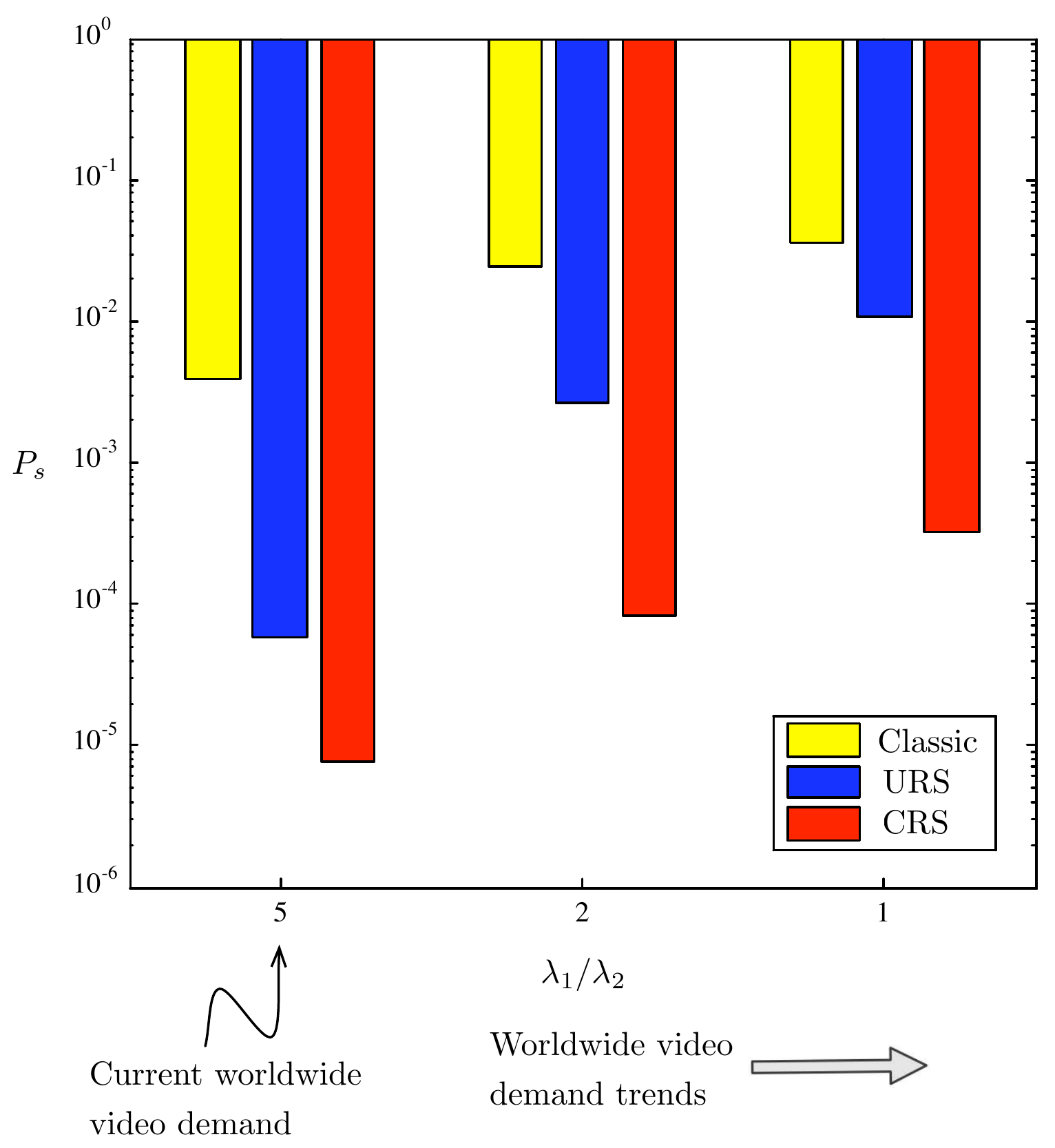}
  \caption{A case study comparing the saturation probability $P_s$ for three different \ac{MR} systems: Classic non-resolution aware systems in which no striping occurs and all file segments are of the same copy are on the same drive; URS, and CRS.  In this study results show that as $\lambda_1/\lambda_2$ decreases, overall saturation probability increases.  The gains of CRS over both URS and classic systems are consistent.  In all cases $ c=1$.  Note that non-resolution aware results are obtained using (\ref{eq:PsR}).}
  \label{fig:PsOptim}
\end{figure}

We now consider the effect of the ratio of Type 1 versus Type 2 requests on saturation probability.  This is an important area of focus for \ac{SAN} designers as drive resources can be fixed.  In particular, we set $m = m_{1} + m_{2}$ constant and analyze $P_{s}$ for  both schemes as a function of $m_{2} / m_{1}$ and $ \lambda_1 / \lambda_2$.   Note that we restrict the range of $ m_2$ in the \ac{URS} system so that  all bandwidth is usable, i.e., $ m_{2}/m_{1} \leq 1$.  (The \ac{CRS} scheme does not have this restriction.)    We assume a fair allocation policy between user types to handle the  multiple type requests in the \ac{MR} scheme.  In particular, we do not allow layer allocation policies that  sacrifice the  saturation probability of one user type to minimize others; a standard quality of service should apply for all user types.  To do so, we find the optimal allocation policy by minimizing a weighted cost function that averages Type 1 and Type 2 requests, given by
\begin{align}
P_b^1 + c P_b^2 
\end{align}
where $P_b^i$ is Type $i$ blocking probability, and $c \in \mathcal{R}$.  Given $ \lambda_1 / \lambda_2$, we then find
\begin{align}
  \underset{m_2}{\min} P_b^1 + c P_b^2 \, .
\end{align}
We then compare the saturation probabilities given these numerically generated optimal \ac{MR} layout strategies.  Results are shown in Fig.~\ref{fig:PsOptim}.  In this result we consider $ \lambda_1 / \lambda_2$ to be equal to 5 as an estimate of current worldwide video-on-demand \ac{HD} versus non-HD trends \cite{Cisco:12}, and note that the trend is increasing in favor  of \ac{HD} traffic.  It was experimentally found that the optimal storage allocation was relatively insensitive to changes in $c$, if changes were kept to the same order of magnitude.  For simplicity, we set $ c=1$ in all plots.  

This plot shows two primary trends.  First, in some cases we can achieve an order of magnitude gain by moving from non-resolution aware to \ac{URS} systems with layers spread across multiple drives.  Second, the \ac{CRS} gain increases as $ \lambda_1/ \lambda_2$ decreases.  This is intuitive because as $ \lambda_1/\lambda_2$ decreases, the required bandwidth to service the greater proportion of multi-layer requests increases.  This implies that the overall system saturation probability will increase and the more saturated the system, the greater the relative gain from increasing scheduling options from \ac{CRS}.  Visually, as $ \lambda_2$ increases, the Markov process average point in Fig.~\ref{fig:MPBoundariesSubFig} moves from the bottom left to the top right hand side of the diagram.  

The \ac{CRS} scheme allows designers and operators greater flexibility in allocating layers and increases focus on quality of video experience through \ac{HD} servicing.  This additional \ac{CRS} flexibility could be a powerful motivator for implementing \ac{CRS} with \ac{MR} codes in future systems.

\section{Discussion}
\label{sec:futWork}

We have explored two types of network coded storage schemes, one for single-resolution and another for multi-resolution storage systems.  

For the \ac{SR} example, we introduced a mapping technique to analyze \acp{SAN} as independent $M/G/K/K$ queues and derived the blocking probability for \ac{UCS} schemes.  These \ac{UCS} schemes were then contrasted with \ac{NCS}.  

The blocking probability gains of \ac{NCS} over \ac{UCS} are dependent on the stripe rate and the Kleinrock independence assumption of arrivals.  The Kleinrock assumption for the independence of incoming chunk read requests requires significant traffic mixing and moderate-to-heavy traffic loads \cite{Kle:B76}.  This models reality most closely in large \acp{SAN} with sufficient traffic-mixing from different users and with traffic loads such as those found in enterprise-level \acp{DC}.  In contrast, in smaller \acp{SAN}, such as those found in closet \acp{DC}, the number of users and the traffic load are smaller and the correlated effects of arrivals between chunks in the same file become more important.  

Given the Kleinrock assumption, results show that blocking probability improvements scale well with striping rates, i.e., as striping rates increase with \ac{NCS}, the number of required file copies decreases.  This may motivate the exploration of very high stripe rates in certain systems for which blocking probability metrics are of particular concern.  Over the last two decades, application bitrate demand growth has continued to outpace increases in drives' I/O growth; although we do not specifically advocate for very high stripe-rates, as long as this trend continues striping  is likely to remain a widespread technique and it may be useful to explore the benefits of very high stripe-rates.

In general, this paper focuses on modeling key \ac{SAN} system components, showing potentially large gains from coding in practical systems.  However, there are numerous additional systems engineering issues that this paper has not considered that would help more precisely quantify these gains.  Such features include dynamic replication of content, and streaming system properties such as time-shifting.  
Additional future work includes extending analysis of \ac{NCS} to correlated arrivals between chunks and to inter-SAN architectures, including \ac{DC} interconnect modeling.

In the \ac{MR} case, a number of areas for future work exist that could build upon the preliminary results obtained in this paper.  It would be insightful to identify analytical bounds between saturation probabilities of the three schemes.  Potential techniques include using mean-field approximations to find the steady state distribution, as well as Brownian motion approximations.  In addition, removing the perfect scheduling assumption would be insightful.  Finally, the \ac{MR} scheme could also be extended for multiple-description codes.

\section{Conclusions}
\label{sec:conc}

In this paper, we have shown that coding can be used in \acp{SAN} to improve various quality of service metrics under normal \ac{SAN} operating conditions, without requiring additional storage space.  For our analysis, we developed a model capturing  modern characteristics such as constrained I/O access bandwidth limitations.   In \ac{SR} storage systems with striping,  the use of \ac{NCS} may be able to provide up to an order of magnitude in blocking probability savings.  In \ac{MR} systems, \ac{CRS} can reduce saturation probability in dynamic networks by up to an order of magnitude.  The spreading of \ac{MR} code layers across drives, coupled with \ac{CRS}, has the potential to yield sizable performance gains in \ac{SAN} systems and it warrants further exploration.  

\bibliographystyle{IEEEtran} \bibliography{IEEEabrv,StringDefinitions,shared,JSAC}

\end{document}